   \documentclass[twocolumn]{emulateapj}
   \usepackage{times}

\shorttitle{C2D IRS I: Silicate emission and grain growth}
\shortauthors{Kessler-Silacci et al.}

\begin{document}


\title{C2D Spitzer-IRS spectra of disks around T~Tauri stars: I. Silicate
emission and grain growth}


\author{Jacqueline Kessler-Silacci\altaffilmark{1}, 
        Jean-Charles Augereau\altaffilmark{2,3},
        Cornelis P. Dullemond\altaffilmark{4}, 
        Vincent Geers\altaffilmark{2}, 
        Fred Lahuis\altaffilmark{2,5}, 
        Neal J. Evans, II\altaffilmark{1},
        Ewine F. van Dishoeck\altaffilmark{2},
        Geoffrey A. Blake\altaffilmark{6}, 
	A. C. Adwin Boogert\altaffilmark{7}, 
        Joanna Brown\altaffilmark{7}, 
        Jes K. J{\o}rgensen\altaffilmark{8},
        Claudia Knez\altaffilmark{1},
        Klaus M. Pontoppidan\altaffilmark{2}
}

\altaffiltext{1}{Department of Astronomy, University of Texas at
Austin, 1 University Station C1400, Austin, TX 78712-0259, USA (jes@astro.as.utexas.edu)}
\altaffiltext{2}{Leiden Observatory, PO Box 9513, 2300 RA Leiden, the
Netherlands} 
\altaffiltext{3}{Laboratoire d'Astrophysique de l'Observatoire de Grenoble, 
B.P. 53, 38041 Grenoble Cedex 9, France}
\altaffiltext{4}{Max-Planck-Institut f{\"u}r Astrophysik, P.O. Box 1317,
D-85741 Garching, Germany}
\altaffiltext{5}{SRON, PO Box 800, 9700 AV Groningen, the Netherlands}
\altaffiltext{6}{Division of GPS, Mail Code 150-21, California
Institute of Technology, Pasadena, CA 91125, USA}
\altaffiltext{7}{Division of PMA, Mail Code 105-24, California
Institute of Technology, Pasadena, CA 91125, USA}
\altaffiltext{8}{Harvard-Smithsonian Center for Astrophysics, 60 Garden
Street, MS42, Cambridge, MA 02138}


\begin{abstract}
Infrared $\sim$5--35~$\mu$m spectra for 40 solar-mass T~Tauri stars and 7 intermediate-mass Herbig Ae stars with circumstellar disks were obtained using the Spitzer Space Telescope as part of the c2d IRS survey. This work complements prior spectroscopic studies of silicate infrared emission from disks, which were focused on intermediate-mass stars, with observations of solar-mass stars limited primarily to the 10~$\mu$m region.  The observed 10 and 20~$\mu$m silicate feature strengths/shapes are consistent with source-to-source variations in grain size. A large fraction of the features are weak and flat, consistent with $\mu$m-sized grains indicating fast grain growth (from 0.1--1.0~$\mu$m in radius).  In addition, approximately half of the T~Tauri star spectra show crystalline silicate features near 28 and 33~$\mu$m indicating significant processing when compared to interstellar grains. A few sources show large 10-to-20~$\mu$m ratios and require even larger grains emitting at 20~$\mu$m than at 10~$\mu$m. This size difference may arise from the difference in the depth into the disk probed by the two silicate emission bands in disks where dust settling has occurred. The 10~$\mu$m feature strength vs. shape trend is not correlated with age or H$\alpha$ equivalent width, suggesting that some amount of turbulent mixing and regeneration of small grains is occurring. The strength vs. shape trend is related to spectral type, however, with M stars showing significantly flatter 10~$\mu$m features (larger grain sizes) than A/B stars.  The connection between spectral type and grain size is interpreted in terms of the variation in the silicate emission radius as a function of stellar luminosity, but could also be indicative of other spectral-type dependent factors (e.g, X-rays, UV radiation, stellar/disk winds, etc.).

\end{abstract}





\section{Introduction}

Dust in disks can be quite different from dust in the interstellar medium 
(ISM).
Observations and chemical modeling (\citealp{grossman_72}, \citealp{gail_98}) suggest that the dust in the early stages of star formation is primarily composed of small ($<$1~$\mu$m) amorphous silicates with strong features at approximately 9.7 and 18.5~$\mu$m \citep[for a summary, see][]{pollack_84}.
Large modifications of the dust occur in the envelopes and disks around young stars, as the initially small grains are processed via collisions and coagulation.  Spectral energy distributions (SED) indicate that grain growth, and 
the corresponding settling of large grains to the disk midplane, is occurring in some disks. Dust settling and growth affect disk temperatures and vertical structures, resulting in dust photospheres that are flatter rather than flared \citep[e.g.,][]{DCHL99,Chiang_v01,DD04a}. 
Additionally, some main sequence stars show evidence of a ``second generation'' of small grains in a debris disk, produced by the collision and fragmentation of planetesimals, that may be quite different from the ``primordial'' dust. 

The spectroscopic study of silicate emission has proven a valuable tracer of grain processing within young circumstellar disks. This method probes small grains via optically thin emission from the surface layer of generally optically thick disks. Studies of $\sim$10~$\mu$m  silicate emission from Herbig Ae/Be stars (hereafter HAEBE) and  T~Tauri stars (hereafter TTs) with disks show early evidence of the growth of these surface-layer grains from $\sim$0.1 to 2.0~$\mu$m \citep[see e.g.,][]{bouwman_01,vanboekel_03}.
Detections of additional spectral features arising from crystalline silicate emission in some HAEBE and TTs disks \citep{waelkens_96,sitko_99,meeus_01,honda_03,acke_v04}, the debris disk $\beta$ Pictoris \citep[e.g.,][]{knacke_f93}, and comets \citep[see review by][]{wooden_02} provide evidence for silicate processing during the disk phase.
Although changes in grain size and composition are closely linked to disk properties and planet formation, the rate and mechanism of grain growth and processing in disks are still not well understood. 

Previous studies of silicate emission from TTs disks were primarily focused on the Si-O stretching mode feature near 10~$\mu$m, which can be observed from the ground.  Grain growth and crystallization, however, have similar effects on the shape of the 10~$\mu$m feature. 
In these previous studies, 
the presence of crystalline silicates was often established through the presence of flux at 11.3~$\mu$m, corresponding to an emission feature of the crystalline Mg-rich silicate forsterite, in addition to the presence of an amorphous olivine feature at 9.8~$\mu$m. The combined effect is a broad, trapezoidal silicate emission feature with a peak near 9.7~$\mu$m and a secondary peak near 11.3~$\mu$m. If crystalline silicates are abundant, the feature can appear broader and flatter. In contrast, growth of amorphous olivines can mimic the effects of crystalline-amorphous silicate mixtures, resulting in weaker, ``flat-topped'' 10~$\mu$m silicate features, with similar flux at 9.8 and 11.3~$\mu$m \citep[e.g.,][]{vanboekel_03, przygodda_03, me_05}.  Indeed, the above studies showed that the flattened silicate features were well matched by models of purely amorphous olivines with grain sizes (radii) of 2~$\mu$m.
Therefore, evaluation of the grain size and/or crystallinity requires observations of a larger spectral region, in which the presence of distinct forsterite or enstatite features (and PAHs, which emit near 7.7, 8.5, and 11.2~$\mu$m) can be assessed.  Complementary studies of the O-Si-O bending mode near 20 $\mu$m probe slightly cooler dust and combined give a better sense of the grain size and crystalline fraction in the disk.

Spectra covering a large spectral region, including several isolated crystalline silicate features (27.5, 33.5, 35.8 and 70~$\mu$m), were obtained for disks around HAEBE stars (and some TTs) with the Infrared Space Observatory (ISO).  ISO observations of several disks around intermediate-mass HAEBE stars clearly established the presence/absence of crystalline silicates \citep[see][for reviews]{MW03,vandishoeck_04}.  Additionally, ISO observations probed the $\sim$10 and $\sim$20~$\mu$m emission features from amorphous silicates in HAEBE disks, allowing grain growth to be studied \citep[see][for a review]{acke_v04}. Due to sensitivity limitations, ISO studies of silicates  focused primarily on intermediate or high-mass stars. Studies of TTs have therefore been limited, for the most part, to the 10~$\mu$m region. Although changes in the shape of the 10~$\mu$m emission features are observed in small samples of TTs \citep{przygodda_03, me_05}, the cause of these changes is unclear and could be a combination of silicate grain growth and crystallization.

In this study, the improved sensitivity of the Infrared Spectrograph (IRS) aboard the Spitzer Space Telescope was used to expand such spectroscopic studies to include a large sample of disks around low-mass, sun-like stars, creating a database analogous to ISO studies of  $\sim$high/intermediate-mass stars.  
A few HAEBE stars were also observed for comparison.  
The data presented here are part of the c2d Spitzer legacy program designed to study the evolution of circumstellar matter `From Molecular Cores to Planet-Forming Disks' \citep{evans_03}.  This paper contains preliminary results from the program and will focus on statistical analysis of the strongest silicate features. It will be followed by a more detailed study of weaker features, including emission from crystalline silicates.  The Spitzer IRS observations are described in $\S$2.  In $\S$3, an inventory of the most prominent silicate emission features will be presented.  The observed spectra will be compared with models of amorphous silicates in $\S$4.  We perform a statistical analysis of the 10 and 20~$\mu$m silicate features and interpret these trends in terms of grain growth in $\S$5. Then in $\S$6, we examine the relationship between the strength-shape trends found in $\S$5 and spectral type, stellar age, and H$\alpha$ equivalent width.  Finally, in $\S$7, we discuss the crystalline silicate emission features in the observed spectra.

\section{Observations}

\begin{deluxetable*}{lllccccccc}
\tablewidth{0pt}
\tabletypesize{\scriptsize}
\tablecaption{Source list \label{tab:obs} }
\tablehead{\colhead{} & \colhead{RA}      & \colhead{DEC}     & \colhead{Observed} & \colhead{AOR} & \colhead{Observation} & \colhead{Age}   & \colhead{H$\alpha$EW\tablenotemark{a}} & \colhead{Spectral} & \colhead{} \\ 
\colhead{Source}      & \colhead{(J2000)} & \colhead{(J2000)} & \colhead{Modules}  & \colhead{Key} & \colhead{Date}        & \colhead{(Myr)} & \colhead{($\rm\AA$)}                    & \colhead{Type}     & \colhead{Ref.}} 
\startdata
RNO 15          		&03 27 47.68  &$+$30 12 04.3   &SL SH LH     &0005633280 &2004-08-30 &...	    &116      &...             &1 		\\
LkH$\alpha$ 327 		&03 33 30.41  &$+$31 10 50.4   &SL SH LH     &0005634560 &2004-02-03 &1.4	    &51--65   &K2              &1,2,3		\\
LkH$\alpha$ 330 		&03 45 48.29  &$+$32 24 11.8   &SL SH LH LL1 &0005634816 &2004-10-25 &...	    &11--20   &G3              &2,3  		\\
IRAS 03446+3254 S		&03 47 47.12  &$+$33 04 03.4   &SL SH LH LL1 &0005635072 &2004-09-29 &...	    &...      &...             &... 		\\
V710 Tau\tablenotemark{b}     	&04 31 57.79  &$+$18 21 36.3   &SH LH        &0005636608 &2004-09-29 &0.7/0.4       &11/89    &M0.5/M3         &4		\\
CoKu Tau 4      		&04 41 16.79  &$+$28 40 00.5   &SH LH        &0005637888 &2004-09-02 &1.2--1.6      &1.8--2.8 &M1.5            &2,5,6  		\\
IRAS 08267-3336 		&08 28 40.70  &$-$33 46 22.3   &SL SH LH LL1 &0005639168 &2004-11-11 &...	    &25--35   &K2--K3          &7,8		\\
SX Cha          		&10 55 59.74  &$-$77 24 39.9   &SH LH        &0005639424 &2004-08-31 &1--4	    &26.7     &M0.5            &9--11 		\\
SY Cha          		&10 56 30.47  &$-$77 11 39.4   &SH LH        &0005639424 &2004-08-31 &2--5	    &24--64   &M0              &9--12   	\\
TW Cha          		&10 59 01.11  &$-$77 22 40.8   &SH LH        &0005639680 &2004-09-01 &20	    &26.1     &M0              &9,11  		\\
VW Cha\tablenotemark{b}         &11 08 00.53  &$-$77 42 28.7   &SH LH        &0005639680 &2004-09-01 &0.4--0.9      &72--147  &K2              &9--12		\\
VZ Cha          		&11 09 23.80  &$-$76 23 20.7   &SH LH        &0005640448 &2004-09-02 &4--10	    &58--71   &K6              &9--12   	\\
WX Cha\tablenotemark{b}        	&11 09 58.75  &$-$77 37 08.9   &SH LH        &0005640192 &2004-09-01 &1--5	    &65.5     &K7--M0          &9--11  		\\
ISO Cha237      		&11 10 11.44  &$-$76 35 29.2   &SH LH        &0005640448 &2004-09-02 &...	    &$<$3     &M0              &13 		\\
C7-11\tablenotemark{b}          &11 10 38.01  &$-$77 32 39.9   &SH LH        &0005640192 &2004-09-01 &0.2--1	    &4.0      &K3              &9,14 		\\
HM 27           		&11 10 49.62  &$-$77 17 51.7   &SH LH        &0005640192 &2004-09-01 &30--40	    &200.0    &K7              &9,11  		\\
XX Cha          		&11 11 39.67  &$-$76 20 15.1   &SH LH        &0005640448 &2004-09-02 &2--40	    &133.5    &M2              &9--11  		\\
T Cha           		&11 57 13.53  &$-$79 21 31.5   &SH LH        &0005641216 &2004-07-18 &$>$12.5	    &2--10.0  &G2--G8          &15--17  	\\
IRAS 12535-7623 		&12 57 11.78  &$-$76 40 11.5   &SH LH        &0011827456 &2004-08-31 &...	    &3.0      &M0              &17,18         	\\
Sz 50           		&13 00 55.37  &$-$77 10 22.2   &SH LH        &0011827456 &2004-03-25 &...	    &46       &M3              &19  		\\
HT Lup\tablenotemark{b}         &15 45 12.87  &$-$34 17 30.6   &SL SH LH     &0009829120 &2004-08-28 &0.4--0.8      &3--7     &K2              &11,12,20 	\\
GW Lup          		&15 46 44.68  &$-$34 30 35.4   &SL SH LH LL1 &0005643520 &2004-08-30 &1.3--3.2      &90--98   &M2--M4          &11,20,21	\\
HM Lup          		&15 47 50.63  &$-$35 28 35.4   &SL LL1 LL2   &0005643776 &2004-08-30 &1.2--2.7      &115--155 &M3              &11,20		\\
Sz 73           		&15 47 56.98  &$-$35 14 35.1   &SL SH LH LL1 &0005644032 &2004-08-30 &2.6--5.4      &97--150  &K2--M0          &11,20,21  	\\
GQ Lup          		&15 49 12.10  &$-$35 39 05.0   &SL SH LH LL1 &0005644032 &2004-08-30 &0.1--0.6      &31--39   &K7--M0          &11,12,20	\\
IM Lup          		&15 56 09.17  &$-$37 56 06.4   &SL SH LH LL1 &0005644800 &2004-08-30 &0.1--0.6      &4.7--8.1 &M0              &11,12,20	\\
RU Lup          		&15 56 42.31  &$-$37 49 15.5   &SL SH LH LL1 &0005644800 &2004-08-30 &0.1--0.5      &136--216 &K7--M0          &11,12,20	\\
RY Lup          		&15 59 28.39  &$-$40 21 51.2   &SL SH LH LL1 &0005644544 &2004-08-30 &1.6--3.2      &7.3      &K0--K4          &12,20		\\
EX Lup          		&16 03 05.52  &$-$40 18 24.9   &SL SH LH LL1 &0005645056 &2004-08-30 &1.4--3.0      &31--43   &M0              &11,12,20	\\
Sz 102          		&16 08 29.70  &$-$39 03 11.3   &SL SH LH     &0009407488 &2004-03-25 &...	    &377      &K0--M4          &20		\\
AS 205\tablenotemark{b}         &16 11 31.35  &$-$18 38 26.1   &SL SH LH     &0005646080 &2004-08-28 &0.1	    &55--155  &K5/M3           &2,3,12,22	\\
VSSG1           		&16 26 18.86  &$-$24 28 19.7   &SH LH        &0005647616 &2004-08-28 &...	    &...      &...             &...		\\
DoAr 24E\tablenotemark{b}       &16 26 23.38  &$-$24 21 00.1   &SH LH        &0005647616 &2003-12-15 &1.5	    &5        &K0              &22,23  		\\
GY23            		&16 26 24.06  &$-$24 24 48.1   &SH LH        &0005647616 &2004-08-28 &...	    &...      &K5--M2          &24		\\
SR 21 N          		&16 27 10.28  &$-$24 19 12.5   &SH LH        &0005647616 &2003-12-15 &1	            &0.54     &G1--G2.5        &22,25		\\
SR 9\tablenotemark{b}           &16 27 40.27  &$-$24 22 04.0   &SH LH        &0012027392 &2004-09-02 &...	    &6--14    &K5--M2          &2,23,24,26	\\
Haro 1-17       		&16 32 21.94  &$-$24 42 14.7   &SL SH LH LL1 &0011827712 &2004-08-29 &...	    &15       &M2.5            &26		\\
RNO 90          		&16 34 09.18  &$-$15 48 16.8   &SL SH LH LL1 &0005650432 &2004-08-28 &6	            &76       &G5              &1		\\
EC 82           		&18 29 56.89  &$+$01 14 46.5   &SL SH LH     &0009407232 &2004-03-27 &...	    &5--11    &M0              &2,27 		\\
CK 4            		&18 29 58.21  &$+$01 15 21.7   &SL SH LH     &0009407232 &2004-03-27 &...	    &...      &...             &...		\\
\\							      					     
\tableline \\						      					     
BF Ori          		&05 37 13.26  &$-$06 35 00.6   &SL SH LH LL1 &0005638144 &2004-10-03 &2.0	    &6--11    &A5--F6          &2,12,28--30     \\
RR Tau          		&05 39 30.52  &$+$26 22 27.0   &SL SH LH LL1 &0005638400 &2004-09-28 &0.6	    &21.2--50 &B8--A5          &2,28,29,31	\\
HD 98922        		&11 22 31.67  &$-$53 22 11.4   &SH LH        &0005640704 &2004-01-04 &...	    &27.9     &B9              &30,31 		\\
DL Cha          		&13 06 08.36  &$-$77 06 27.3   &SH LH        &0005642240 &2004-07-14 &...	    &...      &M6              &22 		\\
HD 135344       		&15 15 48.44  &$-$37 09 16.0   &SH LH        &0005657088 &2004-08-08 &17	    &17.4     &A0--F4          &30,32--34	\\
HD 163296       		&17 56 21.29  &$-$21 57 21.9   &SH LH        &0005650944 &2004-08-28 &4--6	    &12--17   &A0--A2          &15,29--32       \\
VV Ser          		&18 28 47.86  &$+$00 08 39.8   &SL SH LH     &0005651200 &2004-09-01 &0.6	    &22--90   &B1--A3          &3,28--31 	\\
\enddata
\tablecomments{Sources in the top portion of the table are T~Tauri stars; sources in the bottom portion are HAEBE stars.}
\tablenotetext{a} {All H$\alpha$ lines are in emission, except that of SR 21, which is in absorption}
\tablenotetext{b} {This source is a binary in 2MASS K-band images, with a separation $<3''$, and is unresolved with Spitzer IRS.  Spitzer IRS observations are pointed at the center of the binary system. Stellar/disk parameters quoted from the literature include both sources.}
\tablerefs{(1) \citealp{Lev88}; (2) \citealp{CoKu79}; (3) \citealp{FEOM}; (4) \citealp{HSS94}; (5) \citealp{dalessio_05}; (6) \citealp{KBTB}; (7) \citealp{RP93}; (8) \citealp{SS92}; (9) \citealp{LFH96}; (10) \citealp{Hartmann98}; (11) \citealp{AJK}; (12) \citealp{RPL96}; (13) \citealp{Saffe03}; (14) \citealp{Hartigan93}; (15) \citealp{VDA98}; (16) \citealp{AKS95}; (17) \citealp{PDS92}; (18) \citealp{SLD03}; (19) \citealp{HH92}; (20) \citealp{HHKK94}; (21) \citealp{HG89}; (22) \citealp{PGS03}; (23) \citealp{BA92}; (24) \citealp{LR99}; (25) \citealp{MMGC98}; (26) \citealp{Ryd80}; (27) \citealp{GdC97}; (28) \citealp{NGMU97}; (29) \citealp{FM84}; (30) \citealp{Acke05}; (31) \citealp{TdWP94}; (32) \citealp{Thi01}; (33) \citealp{Houk82}; (34) \citealp{Dunkin97}.}
\end{deluxetable*}

Spectra for 40 solar-mass T~Tauri stars and 7 intermediate-mass HAEBE stars 
were obtained using the Infrared Spectrograph (IRS) aboard the Spitzer Space Telescope from December 2003 through December 2004.  The observations and source parameters are described in Table~\ref{tab:obs} for the T~Tauri stars (top) and HAEBE stars (bottom) in our sample.
The spectra were observed with combinations of the short-low (SL), short-high (SH), long-low (LL), and/or long-high (LH) modules (see Table~\ref{tab:obs} column 4).  SL ($\lambda=$ 5.3--14.5~$\mu$m) and LL ($\lambda=$ 14.2--40.0~$\mu$m) spectra have a resolving power of $R=\lambda/\Delta{\lambda}\sim 100$ and the SH ($\lambda=$ 10.0-19.5~$\mu$m) and LH ($\lambda=$ 19.3--37.0~$\mu$m) spectra have a resolving power of $\sim$600.  For approximately half of the sources, SL spectra are part of the GTO programs and not yet available for analysis and, therefore, there is no spectral information short-ward of 10.0~$\mu$m.  Exposure times were chosen to achieve signal-to-noise ratios of ~50 and 100 for sources brighter and fainter than 500 mJy, respectively, with the weakest sources in our sample having fluxes of $\sim$100--200 mJy at 15~$\mu$m.
All sources in the c2d program \citep{evans_03} that were observed prior to December 2004 and show evidence of silicate emission are included (47 sources; see Table~\ref{tab:obs}).

Data reduction was done via the c2d Interactive Analysis (c2dia) reduction
environment. C2dia contains optimized extraction algorithms developed
by the c2d legacy team\footnotemark. 
\footnotetext{The c2d extraction algorithms will become publicly available
through the Spitzer Space Science Center as part of the c2d legacy program.}
For wavelength calibration and IRS aperture definition, tools from
the SMART\footnotemark~ software package \citep{SMART} were used.
\footnotetext{The SMART software package is publicly available on the website 
{\url http://ssc.spitzer.caltech.edu/archanaly/contributed/smart}.}
The extracted spectra were defringed using the IRSFRINGE\footnotemark~
package developed by the c2d team \citep{lahuis_03}.
\footnotetext{IRSFRINGE is included in SMART and available as a
stand-alone package from 
{\url http://ssc.spitzer.caltech.edu/archanaly/contributed/irsfringe/}.}
The spectra were extracted from the Spitzer Science Center (SSC) Basic Calibrated Data (BCD), pipeline version S11.0.2\footnotemark.
\footnotetext{Recently data from the SSC pipeline version S12 as well
as more advanced versions of the extraction algorithms have become
available. We have verified that these do not have a significant 
impact on the reduced spectra and do not affect the results 
presented in this paper.} 
Two different extraction methods were used and compared to reduce 
spectral artifacts resulting from the extraction process.

The first method uses a full aperture extraction for SH and LH and
fixed width aperture extraction for SL and LL. The extraction
aperture was chosen to be large enough to enclose the complete source.
(At the short wavelength end, the extraction aperture 
is wider than that used in the SSC pipeline. However, since the 
SL and LL spectra are corrected for the background, as described below, 
this has a negligable effect on the extracted spectrum.)
Bad pixels are corrected by interpolating in the cross-dispersion direction 
using a fit to the order-averaged source profile. For SL and LL the 
the large apertures allow off-source spectra to be extracted and these 
spectra were used to make wavelength dependent background corrections.  
No background data were available for SH and LH spectra. A spectrum was 
extracted for each position, compared to check for artifacts, and finally 
averaged to produce the final spectrum.

In the second method, all BCDs from both dither positions are combined 
within the extraction algorithm.  The extraction is performed by 
integrating over a source profile fit in the cross-dispersion direction. 
The source profile is a template created from standard star observations 
(including sky measurements for SH and LH).
The width and center of this template is adjusted for each observed 
source in our sample to encompass 95\% of the observed flux.
This fitting process is performed for each source by using the highest  
quality data. Once the source profile is fit, a (uniform) local sky 
background level, which is wavelength dependent, is estimated.
This method also reduces the effects of unidentified bad pixels. 
In particular, in cases where the BCD images are largely affected by 
bad pixels (e.g., for LH), this method quite often gave significant 
improvements over the full aperture extraction.
In the version of c2dia used here, the wavelength and flux calibration 
files produced by SSC pipeline were used. For the S12 version 
of c2dia, the extraction algorithm is further developed and independently 
calibrated using a suite of standard star calibrators.
The extraction algorithms and the calibration involved will be described 
in more detail in Lahuis (in prep.).

\begin{figure*}[t]
\hspace{0.5cm}
\includegraphics[angle=90,width=6.5in]{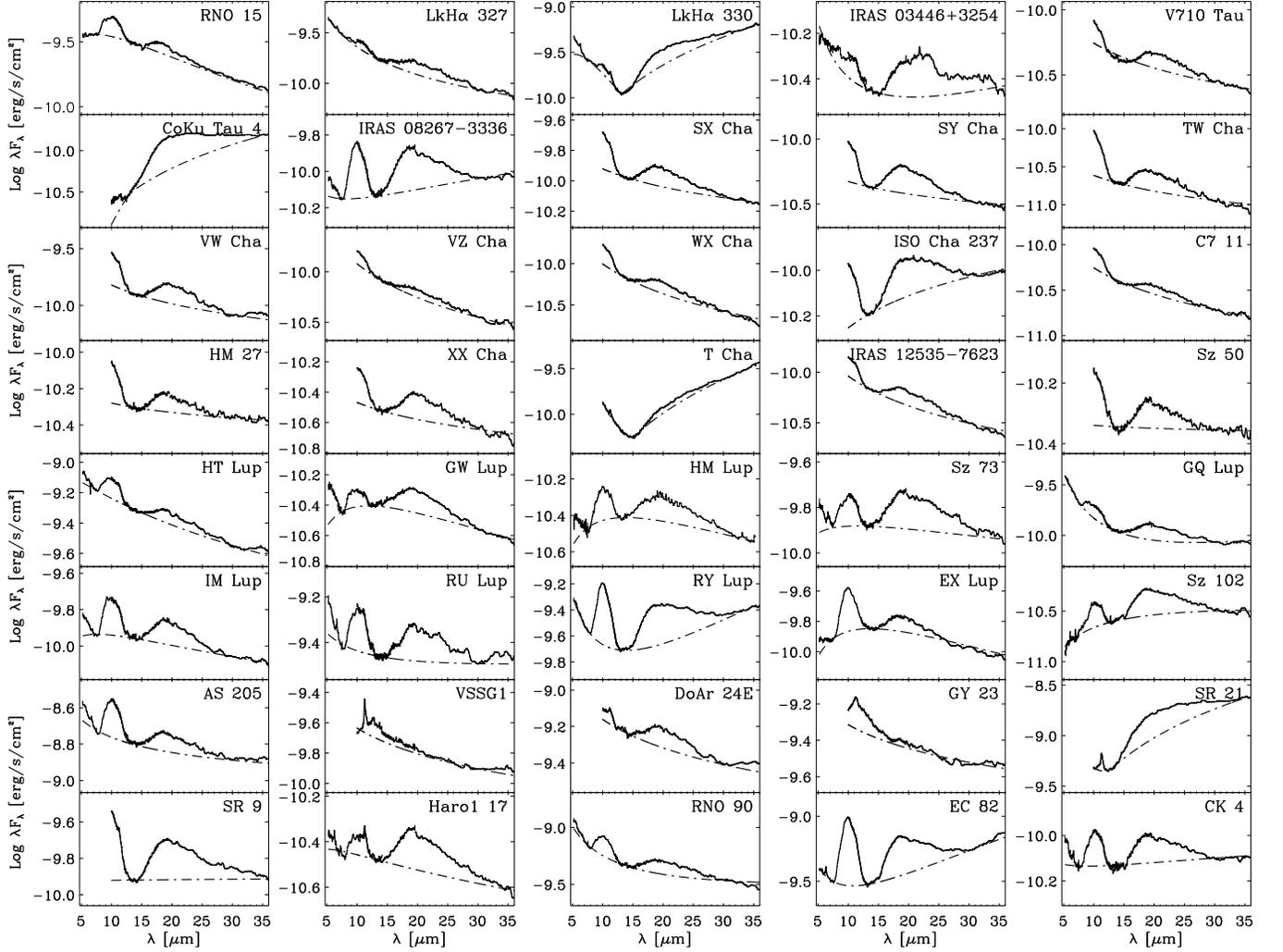}
\vspace{0.5cm}
\figcaption{Spitzer mid- to far-IR spectra for T-Tauri disks. The spectra for each source are shown in units of Log ${\lambda}F_{\lambda}$.  The SH and LH spectra have been median smoothed over 3 and 5 channels, respectively.  No smoothing has been applied to the other modules. The dash-dot line denotes the fit to the continuum described in the text and Table~\ref{tab:features}.  
\label{fig:spec_tt} }
\end{figure*}

To correct for the possibly significant sky contribution in the SH and LH 
spectra, spectra obtained using full aperture extraction were 
corrected using the background estimate from the source profile fitting 
extraction method.
Both extractions were also compared against the spectra extracted from the
SSC pipeline. In a few cases, particular modules suffered from artifacts
due to data selection and/or bad-pixel correction and the corresponding 
modules of the SSC spectrum were used. 
In cases where the SSC spectra were used, a correction to the zero level 
was applied to correct for background emission. Good sections of the
background corrected spectra were used to estimate the zero level
in the other modules. When necessary, individual orders and/or modules 
were multiplied by small ($< 5\%$ and $< 15\%$, respectively) constant 
factors to correct flux offsets between orders/modules.
These flux offsets are likely related to pointing errors
resulting in the source not being centered in the aperture and seem to be
corrected in the SSC pipeline version S12.  

\begin{figure*}
\hspace{0.5cm}
\includegraphics[angle=90,width=7in]{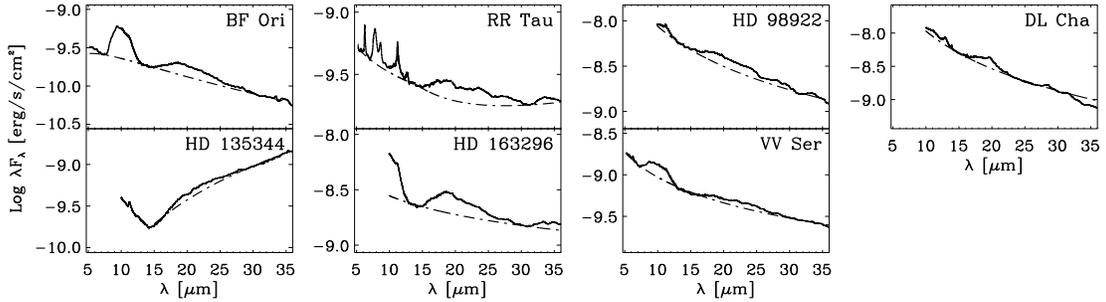}
\vspace{0.5cm}
\figcaption{Spitzer mid to far-IR spectra for HAEBE disks. The spectra for each source are shown in units of Log ${\lambda}F_{\lambda}$.  The SH and LH spectra have been median smoothed over 3 and 5 channels, respectively.  No smoothing has been applied to the other modules. The dash-dot line denotes the fit to the continuum described in the text and Table~\ref{tab:features}.  
\label{fig:spec_h} }
\end{figure*}

\section{Inventory of silicate emission features}
 
\begin{deluxetable*}{lcccccc}
\tablewidth{0pt}
\tabletypesize{\scriptsize}
\tablecaption{Silicate emission features \label{tab:features}}
\tablehead{
\colhead{Species} & \colhead{Silicates\tablenotemark{a}} & \colhead{PAH/Cr.Sil.\tablenotemark{b}} & \colhead{Silicates\tablenotemark{a}} & \colhead{Cr.Sil.\tablenotemark{c}} & \colhead{Cr.Sil.\tablenotemark{d}} & \colhead{Norm.\tablenotemark{e}} \\
\colhead{$\lambda$($\mu$m)} & \colhead{$\sim$10} & \colhead{11.2-11.3} & \colhead{$\sim$20} & \colhead{28-29} & \colhead{33-35} & \colhead{Method} 
}
\startdata
RNO 15          & Y  & T  & T  & Y  & T   &1    \\
LkH$\alpha$ 327 & Y  & Y  & Y  & Y  & Y   &1    \\     
LkH$\alpha$ 330 & Y  & Y  & Y  & T  & N   &3    \\
IRAS 03446+3254 & Y  & Y  & Y  & N  & T   &1    \\ 
V710 Tau        & Y  & T  & Y  & T  & T   &2    \\
CoKu Tau 4      & Y  & T  & Y  & N  & N   &2    \\
IRAS 08267-3336 & Y  & N  & Y  & N  & N   &1    \\
SX Cha          & Y  & T  & Y  & T  & Y   &2    \\
SY Cha          & Y  & T  & Y  & T  & T   &2    \\
TW Cha          & Y  & T  & Y  & T  & T   &2    \\
VW Cha          & Y  & T  & Y  & Y  & Y   &2    \\
VZ Cha          & Y  & T  & T  & Y  & Y   &2    \\
WX Cha          & Y  & T  & Y  & T  & Y   &2    \\
ISO Cha237      & Y  & T  & Y  & T  & T   &2    \\
C7-11           & Y  & T  & Y  & T  & T   &1    \\      
HM 27           & Y  & N  & Y  & T  & T   &2    \\
XX Cha          & Y  & N  & Y  & T  & T   &2    \\
T Cha           & N  & Y  & T  & N  & N   &3    \\
IRAS 12535-7623 & Y  & T  & Y  & Y  & N   &2    \\
Sz 50           & Y  & N  & Y  & N  & T   &2    \\
HT Lup          & Y  & T  & Y  & T  & Y   &1    \\
GW Lup          & Y  & T  & Y  & T  & T   &1    \\
HM Lup          & Y  & T  & T  & T  & T   &1    \\
Sz 73           & Y  & T  & Y  & T  & T   &1    \\
GQ Lup          & Y  & T  & Y  & Y  & Y   &1    \\
IM Lup          & Y  & T  & Y  & T  & T   &1    \\
RU Lup          & Y  & T  & Y  & Y  & Y   &2    \\
RY Lup          & Y  & N  & Y  & N  & N   &1    \\
EX Lup          & Y  & T  & Y  & T  & T   &1    \\
Sz 102          & Y  & T  & Y  & N  & Y   &1    \\
AS 205          & Y  & T  & Y  & T  & T   &1    \\
VSSG1           & Y  & Y  & N  & Y  & T   &2    \\
DoAr 24E        & Y  & Y  & Y  & Y  & T   &2    \\
GY23            & N  & Y  & N  & Y  & T   &2    \\
SR 21           & N  & Y  & Y  & T  & T   &3    \\
SR 9            & Y  & T  & Y  & T  & N   &2    \\
Haro 1-17       & Y  & Y  & Y  & Y  & Y   &1    \\
RNO 90          & Y  & N  & Y  & T  & N   &1    \\
EC 82           & Y  & N  & Y  & N  & T   &1    \\
CK 4            & Y  & T  & Y  & T  & Y   &1    \\
\\
\hline \\    	   	        
BF Ori          & Y  & T  & Y  & T  & Y   &1    \\
RR Tau          & N  & Y  & Y  & Y  & Y   &1    \\
HD 98922        & T  & Y  & Y  & Y  & Y   &2    \\     
DL Cha          & T  & Y  & Y  & Y  & N   &2    \\     
HD 135344       & N  & Y  & T  & N  & N   &3    \\
HD 163296       & Y  & T  & Y  & T  & Y   &2    \\
VV Ser          & Y  & T  & Y  & T  & Y   &1    \\   
\enddata	
\tablecomments{Sources in the top portion of the table are T~Tauri stars; sources in the bottom portion are HAEBE stars.}
\tablenotetext{a} {Broad, primarily amorphous olivine/pyroxene emission features.}					 
\tablenotetext{b} {Feature includes PAH at 11.2~$\mu$m and crystalline forsterite at 11.3~$\mu$m (Geers et al., in prep.).}
\tablenotetext{c} {Feature includes crystalline forsterite near 27.9~$\mu$m and/or crystalline enstatite near 28.5~$\mu$m}
\tablenotetext{d} {Feature includes crystalline forsterite near 33.6~$\mu$m and/or crystalline enstatite near 34.5~$\mu$m}
\tablenotetext{e} {Continuum normalization procedure as described in the text}
\tablecomments{Features are marked with a ``Y'' if detected, ``N'' if not detected, or ``T'' if the identification is tentative.}
\end{deluxetable*}

The extracted spectra of TTs and HAEBE disks are shown in Figure~\ref{fig:spec_tt} and Figure~\ref{fig:spec_h}, respectively.  Two sources, HT Lup and HD 163296, were observed twice, once in IRS-staring mode and once in spectral mapping mode.  The two observations for each source do not differ significantly; integrated intensities, peak-to-continuum fluxes, and feature widths agree to within $\sim$7\% for the 10 and 20~$\mu$m features. Therefore, only one spectrum for each source is included in Figures~\ref{fig:spec_tt} and \ref{fig:spec_h} and Table~\ref{tab:features}.

An inventory of the silicate emission features is presented in Table~\ref{tab:features}.  The features listed are relatively isolated and can be clearly identified in the spectra.  The peak wavelengths of the silicate features depend on the exact composition and grain size of the dust.  Therefore, the broad amorphous silicate features, corresponding to the Si-O stretching and O-Si-O bending modes, are labeled as $\sim$10 and $\sim$20~$\mu$m. The peaks of the identified crystalline silicate features may vary within the wavelength ranges given in Table~\ref{tab:features}, which correspond to the ranges in peak wavelengths for features of crystalline enstatite and forsterite of sizes between 1 and 10~$\mu$m.
Features are marked with ``Y'', ``T'', or ``N'' if the feature is detected, tentatively detected, or absent.  

Emission at 10 and 20~$\mu$m from amorphous silicates dominates the spectra of most of the disks observed.
Broad 20~$\mu$m amorphous silicate features can be seen in almost all of the spectra, while the 10~$\mu$m amorphous silicate features are slightly less prominent (seen in 42 of 47 spectra).  Emission near 11.3~$\mu$m, due to PAH or crystalline forsterite, is clearly identified in $\sim$1/3 of the spectra. This is a lower limit, however, as tentative emission is seen in several additional spectra in which the presence of artifacts prevent further analysis. ISO observations of disks around HAEBE stars \citep{acke_v04} report 11.3~$\mu$m emission from $\sim$1/4 of the observed sources, but the same studies find that the fraction showing PAH emission at 7.7~$\mu$m is much higher (57\%).

\begin{figure}
\includegraphics[angle=90,width=7.5cm]{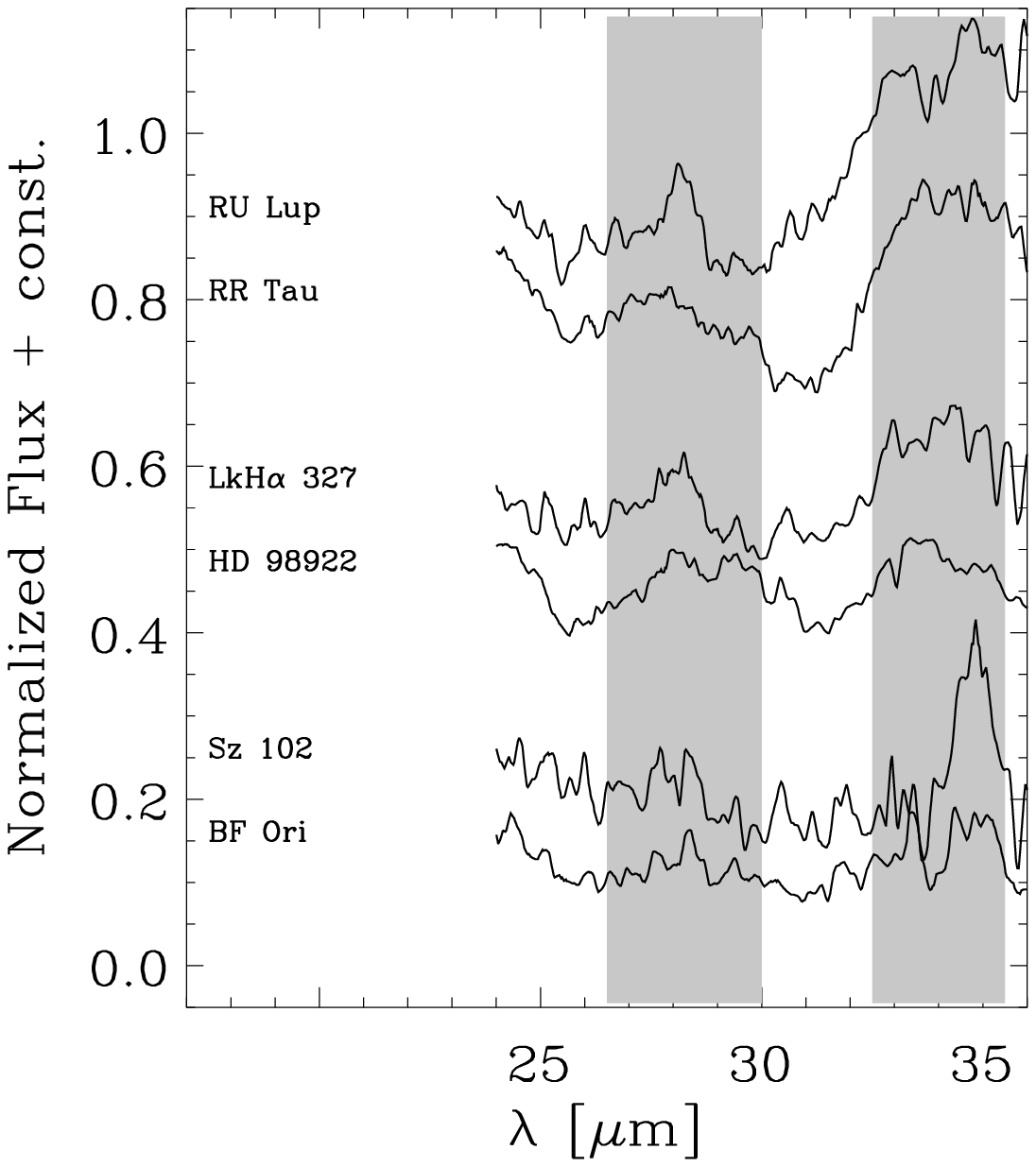}
\figcaption{Continuum normalized and smoothed LH spectra for 6 of the sources that show emission from crystalline silicates in the 26.5--30 and/or 33-35~$\mu$m regions (depicted by shading).  Each pair of spectra contains one HAEBE star (bottom) and one T~Tauri star (top).  Features in the 26.5--30~$\mu$m region tend to be narrower for the T~Tauri stars.
\label{fig:f2533um}}
\end{figure}

The lattice modes of crystalline silicates at $\lambda > 25$~$\mu$m appear to be more prominent than the 11.3~$\mu$m feature in our data and do not suffer from confusion with PAH emission. Approximately half of the disks in our sample show emission features near 28~$\mu$m and in the 33-35~$\mu$m region, characteristic of crystalline enstatite and forsterite. The 24 to 36~$\mu$m region is shown for 6 of these sources in Figure~\ref{fig:f2533um}, with shaded regions depicting the typical positions of enstatite and forsterite features. These data indicate that crystalline silicates can be quite prominent in TTs disks. This is in surprising contrast to ground-based studies of silicates in the 10~$\mu$m region, which found crystalline silicate emission from very few TTs \citep[see][for summary]{me_05}. 

As the 10 and 20~$\mu$m emission features from amorphous silicates are the most easily identified and isolated, the rest of this paper will focus on the analysis of these features. PAHs will be discussed in detail in Geers et al. (in prep.) and SEDs will be modeled in Augereau et al. (in prep.).  Silicate composition will be modeled in more detail by Kessler-Silacci et al. (in prep.), including analysis of all crystalline silicate features.

\section{Comparisons with models of amorphous silicates}

As mentioned above, the peak wavelengths, strengths and shapes of the silicate features depend on the exact composition, sizes, and shapes of the silicates.  Therefore, in order to interpret the observed silicate emission, we must compare our data with models of the emission from a variety of silicate grains.  Modeled grain opacities can be compared to the observed emission once the continuum is removed.  In $\S$4.1, we describe the method of continuum normalization used here.  Our procedure for modeling of grain opacities from laboratory optical constants is described in $\S$4.2.  Finally, in $\S$4.3, we will quantitatively compare modeled and observed opacities in the 10 and 20~$\mu$m regions.

\subsection{Continuum normalization}

Continuum normalization was performed using one of three methods, as noted in Table~\ref{tab:features}, and described as:
\begin{enumerate}
 \item For spectra that include SL (e.g., RNO 15), a 2nd order polynomial was fit to the continuum in the following regions: 6.8--7.5~$\mu$m, 12.5--13.5~$\mu$m, and 30--36~$\mu$m.
 \item For spectra that do not include SL (e.g., V710 Tau), the continuum was obtained with a linear fit to the 12.5--13.5~$\mu$m and 30--36~$\mu$m regions.
 \item In cases where the SED appears to fall sharply with increasing wavelength short-ward of 12--15~$\mu$m and then rises sharply beyond $\sim$15~$\mu$m (e.g., LkH$\alpha$ 330), we fit the 10 and 20~$\mu$m features separately, using a 2nd order polynomial fit for wavelengths short-ward of $\sim$15~$\mu$m and a linear fit long-ward of $\sim$15~$\mu$m.  These sources will be discussed in more detail in Brown et al. (in prep.).
 \end{enumerate}
The continuum fits are overlaid on the observed spectra in Figures~\ref{fig:spec_tt} and \ref{fig:spec_h} as dash-dotted lines.
The continuum fitting regions were chosen to represent the areas least affected by features in the spectra and, as shown, vary slightly from source to source.  Opacities obtained from laboratory spectra of amorphous olivines (see $\S$4) indicate that the regions chosen above should be clear of amorphous silicate emission features, with the exception of the region from $\sim$13--15~$\mu$m, which likely contains contributions from the wings of both the 10 and 20~$\mu$m features.  Thus, the feature strengths obtained from the continuum normalized spectra will be underestimated.

After the continuum fitting was performed, the normalized spectra $S_{\nu}$ were obtained via application of the formula,
$$S_{\nu}=1+\frac{(F_{\nu}-F_{\nu,c})}{<F_{\nu,c}>},$$
 where $F_{\nu}$ is the observed spectrum and $F_{\nu,c}$ is the fitted continuum, both in units of Jy, and $<F_{\nu,c}>$ is the frequency averaged continuum flux.  We divide by a frequency averaged value of the continuum flux in order to remove any dependence on the slope of the continuum. 
This normalization method, including the offset of 1.0, is consistent with the method used by \citet{vanboekel_03} and \citet{przygodda_03}, to which we compare in $\S$5.  
As the continuum may change dramatically between 10 and 20~$\mu$m for some sources (e.g., CoKu Tau 4), $<F_{\nu,c}>$ is calculated separately for regions of $\lambda=5$--13~$\mu$m and $\lambda=13$--37~$\mu$m. 
If $F_{\nu,c}$ is used in place of $<F_{\nu,c}>$, then the normalization would reduce to $F_{\nu}/F_{\nu,c}$, which is approximately equal to the optical depth for optically thin emission;
$$\frac{F_{\nu}}{F_{\nu,c}}=1-{e}^{-\tau_{\nu}} \approx \tau_{\nu} \;\;\; (\tau_{\nu} << 1),$$
where $\tau_{\nu}$ is the optical depth as a function of frequency.
%
Comparison between the two continuum normalization methods for the spectra in the c2d sample shows that the 10 and 20~$\mu$m feature strengths and shapes do not differ significantly ($\sim5$\% and $\sim10$\%, respectively).  

\begin{figure}
\includegraphics[angle=90,height=6.5cm]{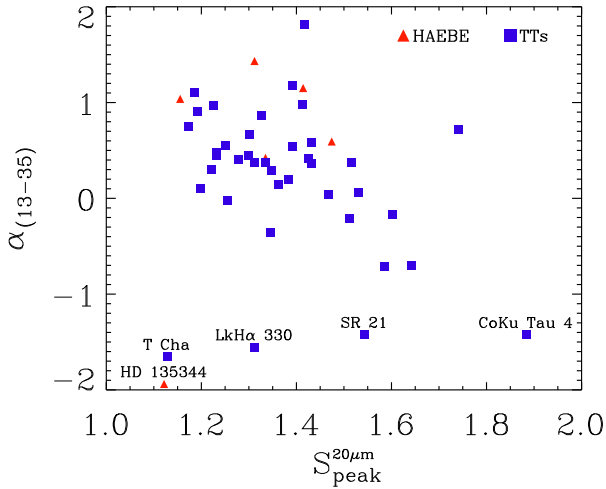}  
\figcaption{Evaluation of the continuum normalization method at 20~$\mu$m.  13-35~$\mu$m spectral indices are plotted versus the strengths of the continuum normalized 20~$\mu$m silicate emission features for all sources.  TTs are denoted by filled squares and HAEBE are denoted by filled triangles.  The mid-IR spectral indices and 20~$\mu$m feature strengths appear to be at most weakly inversely correlated ($r=-0.2$, 80\% significance), with more steeply rising spectra corresponding to stronger 20~$\mu$m silicate emission features.
\label{fig:cont20um}}
\end{figure}

Although continuum normalization has been previously used to interpret ground-based and ISO observations of 10~$\mu$m silicate emission features \citep{honda_03,vanboekel_03,przygodda_03,acke_v04}, it was previously not applied to 20~$\mu$m silicate features because the emission feature strength can be confused by a rising continuum.  To address this issue, we plot the spectral index ($\alpha$) of the derived continuum from 13 to 35~$\mu$m
versus the strength of the continuum normalized 20~$\mu$m feature ($S^{20{\mu}m}_{peak}$) in Figure~\ref{fig:cont20um}. The spectral indices are calculated as,
$$\alpha_{a-b} = - \frac{{\rm Log}\; (\lambda_b F_{\lambda_{b}}) - {\rm Log}\; (\lambda_a F_{\lambda_{a}})}{{\rm Log}\; (\lambda_b) - {\rm Log}\; (\lambda_a)}$$
where $\lambda_a=13$~$\mu$m, $\lambda_b=35$~$\mu$m, and $F_{\lambda_{x}}$ is the flux-density at wavelength $\lambda_x$ in units of erg cm$^{-2}$ s$^{-1}$~$\mu$m$^{-1}$.
The bulk of the spectra are relatively flat or falling in units of $\lambda F_{\lambda}$, with $\alpha=0$--1 and $S^{20{\mu}m}_{peak}=1.1$--1.5 (see Figure~\ref{fig:cont20um}).  The five sources for which we use continuum fitting method $3$ have spectra that are steeply rising from 13 to 35~$\mu$m ($\alpha=-$1 to $-$2) and lie along the bottom of Figure~\ref{fig:cont20um}.  A few other sources with rising spectra (e.g., RY Lup, EC 82) can be seen in the right of Figure~\ref{fig:cont20um}.  We find that the spectral index and the derived strength of the 20~$\mu$m feature is at most weakly correlated (at the 80\% significance level), with the more steeply rising spectra corresponding to the stronger 20~$\mu$m silicate features.  There is thus a small chance that the shape/strength parameters derived for a particular source may be related to the continuum fit.  
As stated above, modeling of individual spectra without continuum normalization will be presented in Kessler-Silacci et al. (in prep.).

\begin{figure}
\includegraphics[angle=0,width=8.5cm]{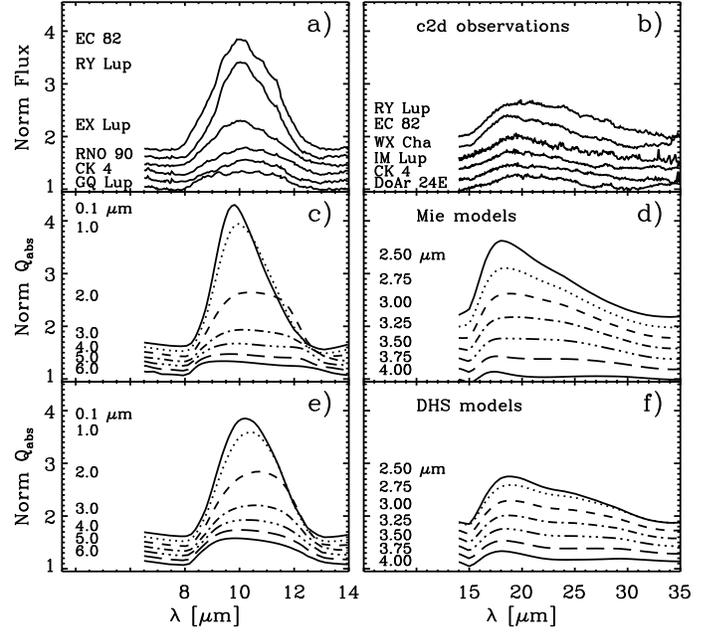} 
\figcaption{Evidence of grain growth in the Si-O stretching and O-Si-O bending mode features. The top panels show the observed normalized spectra in the $a)$ 10~$\mu$m and $b)$ 20~$\mu$m regions for sub-samples of our sources.  The bottom two panels show the normalized absorption efficiencies ($Q_{abs}$) for models of spherical grains of amorphous olivines with various grain sizes calculated for the 10~$\mu$m and 20~$\mu$m regions.  Models of filled homogeneous spheres calculated using Mie theory are shown in $c)$ and $d)$ and models of hollow spheres calculated using DHS theory are shown in $e)$ and $f)$.  Spectra in all panels have been artificially shifted along the y-axis by a constant value as a function of wavelength, such that the spectrum of each source could be seen more clearly.  The minimum of each normalized spectra was 1.0 prior to adding these constants.
\label{fig:4}}
\end{figure}

A representative sample of continuum normalized spectra in the 10 and 20~$\mu$m regions are shown in Figure~\ref{fig:4}a,b.  A large fraction of the observed spectra are similar to GQ Lup, possessing weak, and sometimes flat-topped 10~$\mu$m silicate emission features. A few others possess centrally peaked 10~$\mu$m features, with large peak-to-continuum flux, as is the case for EC 82.  The rest of the sample falls between these two extremes.  Most spectra show 20~$\mu$m features with lower peak-to-continuum flux in the 20~$\mu$m region with respect to that at 10~$\mu$m.

\subsection{Modeling dust optical properties}

\begin{deluxetable}{lcc}
\tablewidth{0pt}
\tabletypesize{\scriptsize}
\tablecaption{List of adopted grain material \label{tab:models} }
\tablehead{\colhead{Dust type} & \colhead{Formula/Name} & \colhead{Refs}}
\startdata
amorphous olivine       &MgSiO$_4$  			&\citet{Dorschner95}  \\ 
amorphous pyroxene      &Mg$_{0.5}$Fe$_{0.5}$SiO$_3$	&\citet{Dorschner95}  \\
crystalline forsterite  &Mg$_2$SiO$_4$     		&\citet{Servoin73}    \\ 
crystalline enstatite   &MgSiO$_3$     			&\citet{Jager98}      \\
amorphous carbon	&ACAR				&\citet{Zubko96}      \\
\enddata
\end{deluxetable}

To aid in the interpretation of the observed silicate emission features, we model the opacities for a sample of grains of different shape, size and composition.  Optical constants are obtained from laboratory data available from the Jena -- St. Petersburg database\footnotemark \footnotetext{http://www.astro.uni-jena.de/Laboratory/Database/databases.html} \citep[][see Table~\ref{tab:models}]{jenadb}.  Absorption efficiencies ($Q_{abs}$) are then calculated for spherical, homogeneous grains of different sizes, representing compact grains, using Mie theory \citep{mie_1908}.  Interstellar dust, however, is not likely to be spherical and images of interplanetary dust particles \citep[see ][for review]{Bradley_03}  suggest that nebular dust likely consists of porous and irregularly shaped aggregates. Additionally, the positions of the features in the mass absorption coefficients of crystalline silicates derived via Mie theory do not agree with observations of astronomical silicates \citep[e.g.,][]{bouwman_01}.  Therefore, to simulate porous or irregularly shaped particles, we calculate opacities for continuous distributions of hollow spheres (DHS), using the statistical method described in \citet{min_05}. In the Mie method, the particle can be represented by a homogeneous, filled sphere.  In the DHS method, a distribution of hollow spherical silicate shells is used such that the vacuum volume filling fractions vary, but the masses remain the same as that of the homogeneous, filled sphere.  The DHS method averages the scattering and absorption/emission cross sections of the set of hollow spheres and these cross sections can be compared to those of the homogeneous filled sphere of the same mass.
The optical properties of individual large hollow spheres are calculated using the algorithm proposed by \citet{TA81} for coated spheres, which was found to be extremely accurate (to 4 decimal places) when compared to measured efficiencies. We apply a simple analytical application of the Rayleigh approximation as described in \citet{min_03} for small grains with $x=2{\pi}a/{\lambda}<<1$ and $|mx|<<1$, where $a$ is the radius of the particle and $m$ is the refractive index.

\begin{figure}
\includegraphics[angle=90,width=8cm]{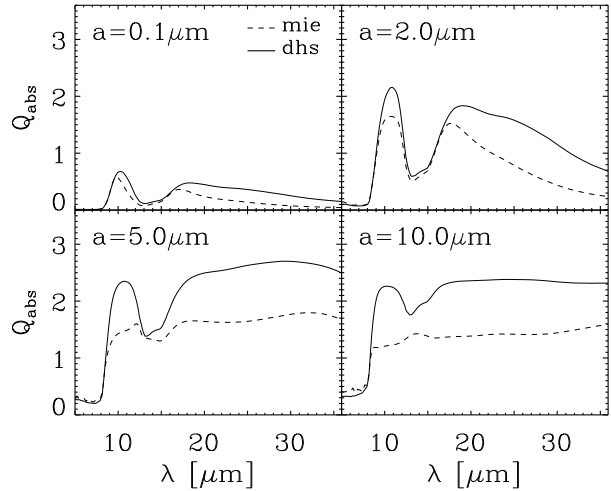}
\figcaption{Absorption efficiencies for amorphous olivine for grain sizes of 0.1, 2, 5, and 10~$\mu$m.  In the upper left panel, $Q_{abs}$ for $a=0.1 \mu$m has been multiplied by a factor of 5. The dashed lines show $Q_{abs}$ calculated using Mie theory for spherical grains. The solid lines show $Q_{abs}$ calculated using the distribution of hollow spheres (DHS) method described in the text. \label{fig:qabsoliv}}
\end{figure}

\begin{figure}
\includegraphics[angle=90,width=8cm]{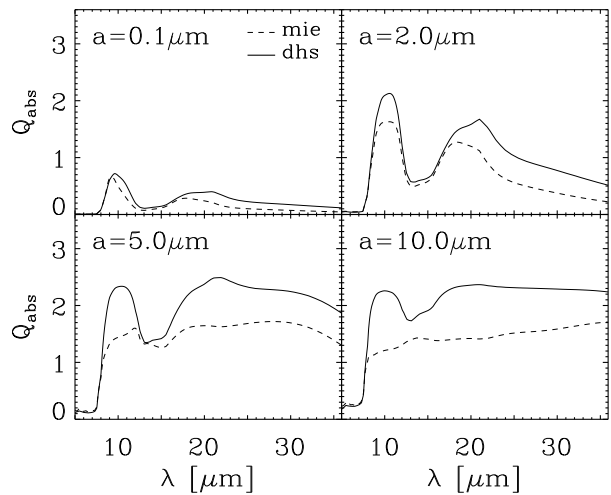}
\figcaption{Absorption efficiencies for amorphous pyroxene for grain sizes of 0.1, 2, 5, and 10~$\mu$m.  In the upper left panel, $Q_{abs}$ for $a=0.1 \mu$m has been multiplied by a factor of 5. Lines are as in Figure~\ref{fig:qabsoliv}. \label{fig:qabspyrox}}
\end{figure}

\begin{figure}
\includegraphics[angle=90,width=8cm]{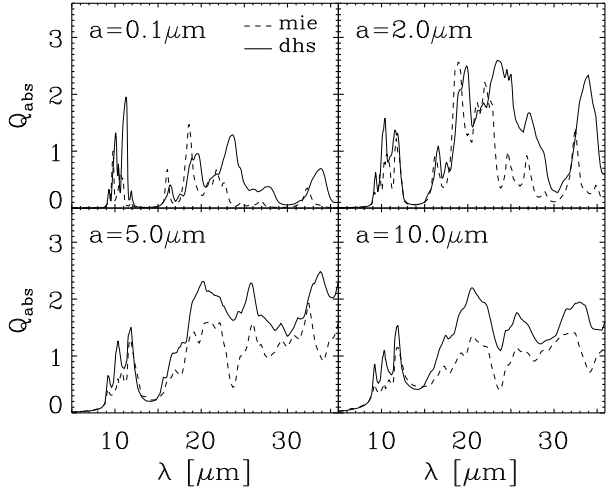}
\figcaption{Absorption efficiencies for crystalline forsterite for grain sizes of 0.1, 2, 5, and 10~$\mu$m.  In the upper left panel, $Q_{abs}$ for $a=0.1 \mu$m has been multiplied by a factor of 5. Lines are as in Figure~\ref{fig:qabsoliv}.  Note that the results of Mie/DHS theory for large ($a>2$ $\mu$m) forsterite grains have not been confirmed by laboratory studies. Additionally, emission feature positions are offset for Mie and DHS models. DHS models have been found to more closely agree with astronomical observations of silicates in disks \citep{vanboekel_05}. \label{fig:qabsfors}}
\end{figure}

Figures~\ref{fig:qabsoliv}--\ref{fig:qabsfors} show the absorption efficiencies ($Q_{abs}$) for grains of amorphous olivine, pyroxene, and crystalline forsterite for a range of sizes between 0.1 and 10~$\mu$m.  Here we have used the grain radius ($a$) of the filled, homogeneous sphere as a proxy for the grain mass, which is held constant for both Mie and DHS models. 
As indicated in Figure~\ref{fig:qabsoliv}, the slope of the overall spectrum varies significantly as a function of grain size.  Thus, in order to compare the model $Q_{abs}$ to the normalized Spitzer spectra, we use a ``continuum'' fitting method which is similar to that applied to the Spitzer spectra. Although this method will most likely not find the physical continuum level, treating the observed and modeled spectra in the same manner will enable a more robust comparison.  Normalized $Q_{abs}$ in the 10 and 20~$\mu$m regions for the modeled grain opacities of amorphous olivines calculated using Mie and DHS are shown in Figure~\ref{fig:4}c,d and Figure~\ref{fig:4}e,f, respectively.  
The comparison of these normalized absorption efficiencies to the observed spectra involves an implicit assumption about the dominance of silicate emission in the 8--35~$\mu$m region.  Disk models \citep[e.g.,][]{pollack_94} based on solar elemental abundances find that the main refractory components are olivine, pyroxene, carbonaceous material and/or organics, water ice, troilite (FeS) and metallic iron.  They find that silicates clearly dominate the opacity for $T>500$ K, but at lower temperatures the shapes and strengths of the 10 and 20~$\mu$m features can be affected by the opacities of organics and water ice.  
The silicate emission studied here arises from the warm surface layers of these disks and ices are unlikely to play any role. Additionally, the excellent correspondence between the modeled silicate opacities and the observed spectra (cf. Figure~\ref{fig:4}), in both contrast and spectral shape, suggest that we are in a regime in which this comparison is valid.

\subsection{Grain growth}

Both the shape and strength of the silicate features are dependent upon grain size, with Mie and DHS models possessing slightly different silicate features. In this section we will compare the 10 and 20~$\mu$m silicate features in the observed spectra to the Mie and DHS models (Figure~\ref{fig:4}).

Both Mie and DHS models show that the 10~$\mu$m silicate feature strength decreases, and the feature flattens, as a function of grain size.
The 10~$\mu$m feature strength for Mie models (Figure~\ref{fig:4}c) is largest for grain sizes of $a\approx0.1$~$\mu$m. When $a=8$~$\mu$m, the feature has almost disappeared.  A similar trend can be seen in the models of hollow spheres using DHS methods in Figure~\ref{fig:4}e.  The shape of the 10~$\mu$m feature is slightly more round for the DHS than Mie models, however, and the peak strength is underestimated (with respect to Mie models) for smaller grain sizes ($a=0.1$~$\mu$m) and overestimated for larger grain sizes ($a=6.0$~$\mu$m).

For the 20~$\mu$m feature (Figure~\ref{fig:4}d,f), again we see that for both Mie and DHS models, the feature strength decreases and broadens when the grain size is increased from 2.5 to 5.0~$\mu$m. The emission feature also appears to shift to longer wavelength with increasing grain size, peaking near 18~$\mu$m (similar to our observations) for Mie and DHS models of amorphous olivines with grain sizes of $a=$ 3--5~$\mu$m.  The shapes of the 20~$\mu$m features of the Mie and DHS models are also slightly different, with DHS models showing a much flatter feature. Additionally, the peak strengths of the DHS models are smaller than those of the Mie models for smaller grain sizes ($a=2.5$~$\mu$m) and larger than the Mie models for larger grain sizes ($a=4.0$~$\mu$m).

The observed spectra (Figure~\ref{fig:4}a,b) are quite similar to the models, differing slightly for the weakest emission features.  The weakest 10~$\mu$m features observed toward several sources are similar to the models of 5~$\mu$m amorphous olivine grains, but appear to be slightly narrower and more sharply peaked at the edges of the feature.  Additional emission from crystalline silicates may be necessary to reproduce the ``boxy'' structure of such features.  The strongest observed 20~$\mu$m features are similar in strength to models of amorphous olivine with grain sizes $>$2~$\mu$m.  The weakest features are similar in strength to models of amorphous olivine with sizes less than approximately 4.5~$\mu$m.  

Thus, the comparison of the observed spectra with modeled amorphous olivine opacities provides a qualitative understanding of the effect of grain growth on the 10 and 20~$\mu$m features observed for our sample. In the following section, we will perform a more quantitative analysis.

\section{Statistical analysis of the 10 and 20~$\mu$m features}

Previous observations of disks around HAEBE and T~Tauri stars have shown that plots of the shape vs. strength of the 10~$\mu$m emission feature may be indicative of grain growth.  In $\S$5.1 we interpret variations in the shape and strength of the observed silicate features in terms of source-to-source variations in grain size. Because it is important to understand the uniqueness of this interpretation, in $\S$5.2 we evaluate the dependence of the 10 and 20~$\mu$m feature strengths on several other grain properties, including the grain size distribution, relative sizes of grains emitting at 10 and 20~$\mu$m, grain composition, crystallinity, and porosity.  

\subsection{Strength-shape trends}

\begin{deluxetable*}{lllccccc}
\tablewidth{0pt}
\tabletypesize{\scriptsize}
\tablecaption{Stellar/Disk parameters for MIR spectra from the literature \label{tab:lit} }
\tablehead{\colhead{} & \colhead{RA\tablenotemark{a}}      & \colhead{DEC\tablenotemark{a}}     & \colhead{Spect.\tablenotemark{b}} & \colhead{Age}   & \colhead{H$\alpha$EW\tablenotemark{c}} & \colhead{Spectral} & \colhead{} \\ 
\colhead{Source}      & \colhead{(J2000)}                   & \colhead{(J2000)}                   & \colhead{Ref.}   & \colhead{(Myr)} & \colhead{($\rm\AA$)}                   & \colhead{Type}     & \colhead{Ref.}} 
\startdata
GG Tau\tablenotemark{d}		&04 32 30.31 &$+$17 31 41.0		&Prz		&1.7	  &31--54    &K6--K7    &1,2,3		\\
AA Tau				&04 34 55.45 &$+$24 28 53.7		&K-S		&0.9--2.4 &37.1	     &K7        &1,2,4,5	\\
Lkca 15				&04 39 17.80 &$+$22 21 04.5		&K-S		&2--11    &23.1	     &K5        &1,4,6		\\
DR Tau				&04 47 05.48 &$+$16 58 42.1		&Prz		&3.8	  &95.4      &K4        &1,2		\\
GM Aur				&04 55 10.90 &$+$30 22 01.0		&K-S		&0.9--1.8 &96.5      &K7        &1,2,4		\\
SU Aur				&04 55 59.38 &$+$30 34 01.5		&Prz		&3.0	  &3.5	     &G2        &2,7		\\
GW Ori				&05 29 08.39 &$+$11 52 12.7		&Prz		&...   	  &27--29    &G5        &2,3		\\
CR Cha				&10 59 06.97 &$-$77 01 40.3		&Prz		&...   	  &30--44    &K2        &3,8		\\
TW Hya				&11 01 51.91 &$-$34 42 17.0		&Prz		&9--10    &194       &K7        &1,3,9,10	\\
Glass I\tablenotemark{d}	&11 08 15.41 &$-$77 33 53.5		&Prz		&2.5	  &22/1.4    &K4        &11--14		\\
WW Cha				&11 10 00.7  &$-$76 34 59		&Prz		&0.3--0.8 &66--67    &K5        &3,4,8,11	\\
Hen 3-600 A			&11 10 28.86 &$-$37 32 04.8		&K-S		&10	  &12.5	     &M4        &13,14		\\
IRAS 14050-4109 		&14 08 10.3  &$-$41 23 53		&K-S		&...   	  &2.0	     &K5        &13,15		\\
RU Lup				&15 56 42.31 &$-$37 49 15.5		&Prz,here	&0.1--0.5 &136--216  &K7--M0    &3,8,16		\\
AS 205 NE			&16 11 31.40 &$-$18 38 24.5		&Prz,here	&0.1	  &154.6     &K5        &2,17		\\
AS 205 SW			&16 11 31.40 &$-$18 38 24.5		&Prz,here	&$<$0.1	  &54.9      &K5        &2,17		\\
DoAr 24E\tablenotemark{d}	&16 26 17.06 &$-$24 20 21.6		&Prz,here	&1.5	  &5         &K0        &17,18		\\
Haro 1-16			&16 31 33.53 &$-$24 27 33.4		&Prz		&...   	  &59--76    &K2--K3    &2,8,18		\\
AK Sco				&16 54 44.85 &$-$36 53 18.6		&Prz		&1.0	  &3--9	     &F5        &3,13,19	\\
S CrA NW			&19 01 08.60 &$-$36 57 20.0		&Prz		&3.0      &73.       &K3        &13,17		\\
S CrA SE			&19 01 08.60 &$-$36 57 20.0		&Prz		&1.0	  &61        &MO        &13,17		\\
\\
\hline \\
MWC 480				&04 58 46.27 &$+$29 50 37.0		&K-S		&5--6     &18.3	     &A2--A3    &1,20--23	\\
UX Ori A			&05 04 29.99 &$-$03 47 14.3		&vB		&3--5     &20.0      &A2--A3    &1,19,20--23	\\
HD 37357			&05 37 47.08 &$-$06 42 30.2		&vB		&...      &...       &A0        &22		\\
HD 37806			&05 41 02.29 &$-$02 43 00.7		&vB		&6.3      &...       &B9--A2    &22--24		\\
HD 95881			&11 01 57.62 &$-$71 30 48.4		&vB		&...      &21.1      &A1--A2    &21,22		\\
HD 98922			&11 22 31.67 &$-$53 22 11.5		&vB,here	&...      &27.9      &B9        &21,22		\\
HD 101412			&11 39 44.46 &$-$60 10 27.7		&vB		&...      &14-20     &B9.5      &13,21,22	\\
HD 104237			&12 00 05.08 &$-$78 11 34.6		&vB		&2        &24.3      &A4        &13,20--23,25	\\
HD 142666			&15 56 40.02 &$-$22 01 40.0		&vB		&10	  &0.8--3    &A7--A8    &13,20--22	\\
HD 144432			&16 06 57.96 &$-$27 43 09.8		&vB		&...      &5--9      &A7--F0    &13,21--24	\\
HD 150193			&16 40 17.92 &$-$23 53 45.2		&vB		&3--5     &5         &A0--A4    &20--22,25	\\
HD 163296			&17 56 21.29 &$-$21 57 21.9		&vB,K-S,here	&5        &14.5      &A0--A2    &1,21--23,26	\\
HD 179218			&19 11 11.25 &$+$15 47 15.6		&K-S		&0.1	  &18.2	     &B9--A0    &21--23		\\
WW Vul				&19 25 58.75 &$+$21 12 31.3		&K-S		&$>$10    &18--24    &B9--A3    &3,4,20--22	\\
HD 184761			&19 34 58.97 &$+$27 13 31.2		&K-S		&...   	  &...	     &A8        &27		\\
\enddata
\tablecomments{Sources in the top portion of the table are T~Tauri stars; sources in the bottom portion are HAEBE stars.}
\tablenotetext{a} {The RA and DEC quoted here were obtained from SIMBAD and do not necessarily represent the position of the referenced MIR spectroscopic observations.}
\tablenotetext{b} {References for silicate spectroscopy: vB $=$ \citealp{vanboekel_03}; Prz $=$ \citealp{przygodda_03}; K-S $=$ \citealp{me_05}; here $=$ this paper.}
\tablenotetext{c} {All H$\alpha$ lines are in emission}
\tablenotetext{d} {This source is a binary that is unresolved in the referenced MIR spectroscopic observations.  Stellar/disk parameters quoted from the literature include both sources.}
\tablerefs {(1) \citealp{Thi01}; (2) \citealp{CoKu79}; (3) \citealp{RPL96}; (4) \citealp{Hartmann98}; (5) \citealp{Strom89}; (6) \citealp{Poncet98}; (7) \citealp{Lev88}; (8) \citealp{AJK}; (9) \citealp{Webb99}; (10) \citealp{Torres03}; (11) \citealp{LFH96}; (12) \citealp{GaS}; (13) \citealp{PDS92}; (14) \citealp{CGMH}; (15) \citealp{G-HH02}; (16) \citealp{HHKK94}; (17) \citealp{PGS03}; (18) \citealp{BA92}; (19) \citealp{HBC}; (20) \citealp{NGMU97};  (21) \citealp{Acke05}; (22) \citealp{TdWP94}; (23) \citealp{VDA98}; (24) \citealp{meeus_01}; (25) \citealp{VDA97}; (26) \citealp{FM84}; (27) \citealp{Miroshnichenko99}.}
\end{deluxetable*}

\begin{figure*}
\includegraphics[angle=90,width=8.85cm]{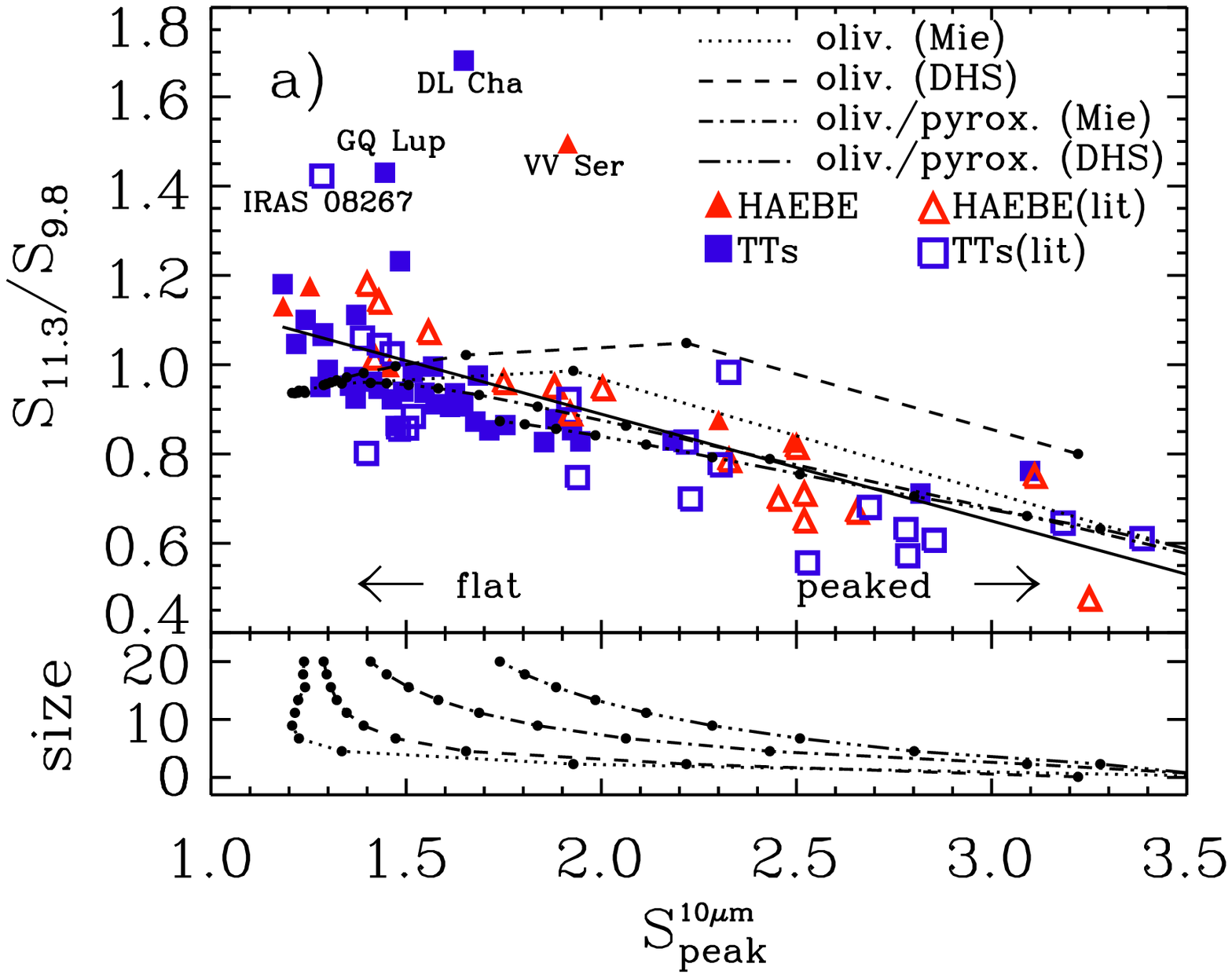}
\hspace{0.2cm}
\includegraphics[angle=90,width=8.85cm]{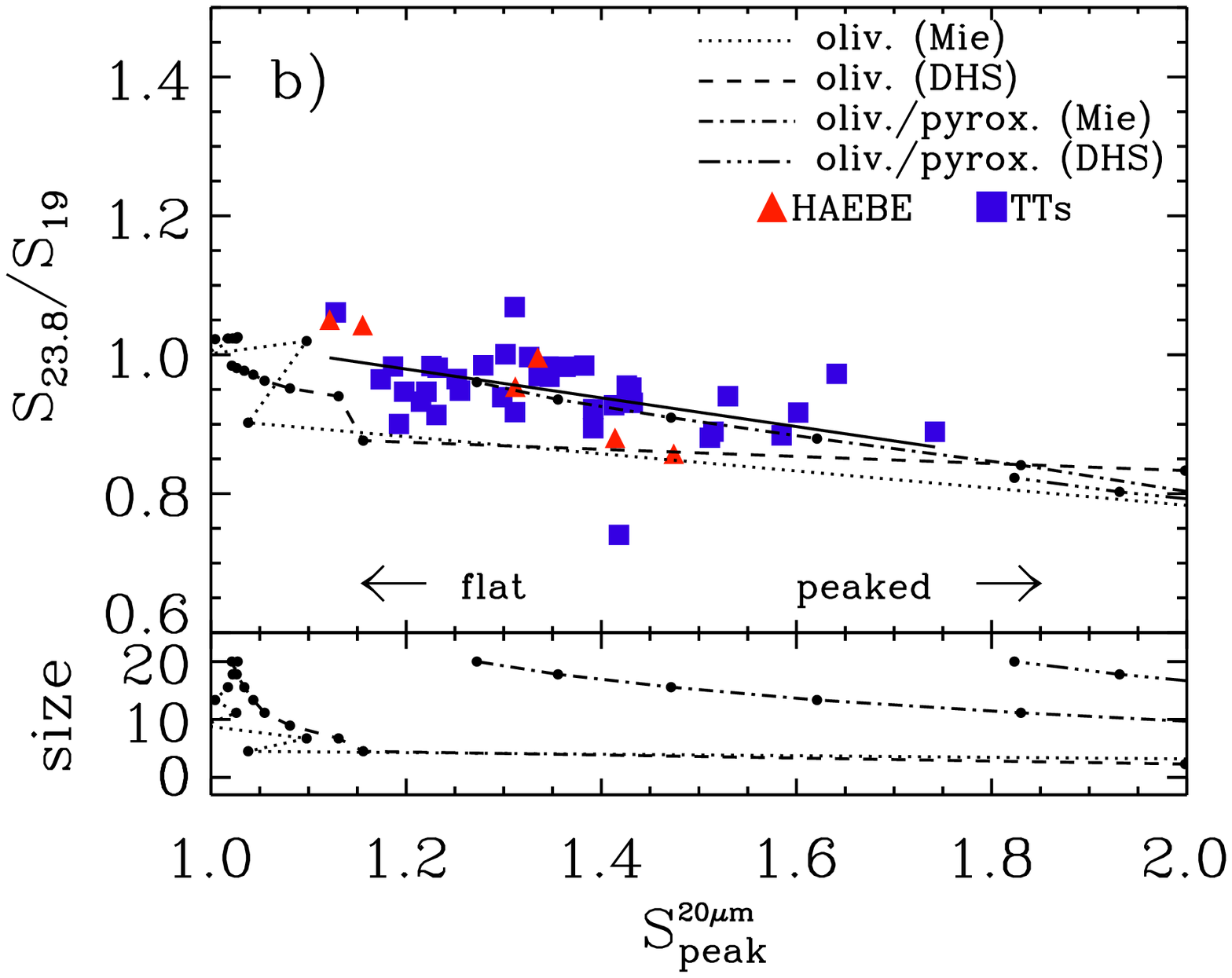}
\figcaption{Shape and strength of the $a)$ 10~$\mu$m and $b)$ 20~$\mu$m silicate features. The shapes of the features (probed by the normalized flux ratios, $S_{11.3}/S_{9.8}$ and $S_{23.8}/S_{19}$) are plotted versus the feature strengths ($S^{10{\mu}m}_{peak}$ and $S^{20{\mu}m}_{peak}$) in the top panels of each figure.  The data observed in this study are denoted by filled squares (TTs) and filled triangles (HAEBE). Data for 10~$\mu$m silicate features from the literature \citep{vanboekel_03, przygodda_03,me_05} are always represented by open symbols. The black line in $a)$ indicates the correlation found for the 10~$\mu$m features including all data, and follows the relation, $y=A+Bx$, where $A=1.37\pm0.05$ and $B=-0.24\pm0.03$, with a correlation coefficient of $r=-0.7$.  The 20~$\mu$m features also show a correlation ($r=-0.5$.), with $A=1.23\pm0.08$ and $B=-0.21\pm0.06$. Models of amorphous olivines, pyroxenes and mixtures of the two for various grain shapes are overlaid as dashed-dotted lines connecting points representing grains with a linear size distribution between 0.1 and 20~$\mu$m (see text for details).  The bottom panels of each figure show the grain size vs. peak strength for the models.
\label{fig:vb}}
\end{figure*}

We find that the shapes of the continuum normalized 10~$\mu$m emission feature from HAEBE and TTs are proportional to the feature strengths (as was previous noted by \citealp{vanboekel_03}, and later \citealp{przygodda_03} and \citealp{me_05}). In Figure~\ref{fig:vb}a, we plot the shape of the 10~$\mu$m feature versus the feature strength, $S^{10{\mu}m}_{peak}$. As a proxy for the feature shape, we use the ratio of the normalized flux at the peak of the crystalline silicate feature relative to that of the amorphous silicate feature ($S_{11.3}/S_{9.8}$).  The fluxes are integrated over regions of $\pm$0.1~$\mu$m around the central wavelength. Data from the literature ($lit$, \citealp{vanboekel_03,przygodda_03,me_05} open symbols) are plotted in addition to the data collected in this study ($c2d$, filled symbols).  Source parameters and original references for the $lit$ sample are listed in Table~\ref{tab:lit}. There are five sources that overlap between the {\it c2d} and {\it lit} samples: RU Lup, AS 205, DoAr 24E, HD 98922, and HD 163296.  The 10~$\mu$m feature strengths and shapes of these sources are consistent to within $11\pm7$\%.  In the statistical analysis, we use only our data for the duplicate sources, except in the case of AS 205 for which we will use the \citet{przygodda_03} spectra for each source in the binary. There is a strong correlation for the entire sample of 10~$\mu$m features, which is consistent with that first noted by \citet{vanboekel_03}. This trend is similar to changes in silicate emission due to grain size variations illustrated in Figure~\ref{fig:4}, with flatter, weaker 10~$\mu$m emission features, characteristic of larger grain sizes, appearing in the upper left of Figure~\ref{fig:vb}a. 
 
In order to interpret this trend more quantitatively, we perform the same analysis for the continuum normalized model opacities discussed in $\S$4.2. We include grains of  amorphous olivine, pyroxene, and mixtures of the two for grain sizes between 0.1 and 20~$\mu$m.  The Maxwell--Garnett mixing rule \citep{MG1904} is used to calculate effective cross sections for all mixtures. 
The model data are overlaid on the observational data in Figure~\ref{fig:vb}a, with lines connecting points of varying grain size for each model.   Note that the peak strength corresponds to different sizes depending on the dust composition and whether DHS or Mie models are used (see bottom panel of Figure~\ref{fig:vb}a).  

The observed trend is best fit with models of amorphous pyroxene/olivine mixtures.  Models of pure amorphous olivines calculated using DHS are inconsistent with the observations.  These models possess generally larger $S_{11.3}$-to-$S_{9.8}$ ratios than the data, particularly for strong features.  Mie models of pure amorphous olivines fit the data much better, but begin to deviate from observations near $S^{10{\mu}m}_{peak}\approx2$. The slope of the observed trend can be best matched using mixtures with 30$\%$ amorphous olivines and 70$\%$ amorphous pyroxenes.  This pyroxene fraction is much larger than that inferred for diffuse ISM grains in the galactic center \citep[15.1\% pyroxene; ][]{kemper_04}, indicating a substantial conversion of olivines to pyroxenes in young stellar environments.  Enhanced pyroxene-to-olivine abundances were previously observed toward high-mass protostars with ISO and may be explained by He$^+$ sputtering in high-velocity shocks \citep[][and references therein]{demyk_01}.

Models of olivine-pyroxene mixtures fit the data best for homogeneous filled spheres (Mie) and hollow spheres (DHS), but do not fit equally well for all grain sizes (peak strengths).  
For the weakest 10~$\mu$m features ($S^{10{\mu}m}_{peak}\le1.75$), hollow sphere models indicate grain sizes of up to 1.3 times those of homogeneous spheres for features of the same strength.
Both mixture models indicate that the largest changes in the shape of the feature occur over a small range of grain sizes, from approximately 1--3~$\mu$m, with exact values depending on the model used. For the weakest silicate features ($S^{10{\mu}m}_{peak}\le1.7$), large changes in grain size result in only small variations in the strength/shape of the silicate feature.  Therefore, it becomes increasingly difficult to determine exact grain sizes for the weakest silicate features.

The data discussed above cannot be explained purely by size variations, and in some cases can indeed be attributed to crystalline silicates and/or PAH emission. For the largest grain sizes, Figure~\ref{fig:vb}a shows several spectra with 11.3-to-9.8~$\mu$m ratios that consistently lie above the modeled ratios.  These data may indicate the presence of crystalline forsterite or PAH emission in the 11.2--11.3~$\mu$m region.  For a cutoff at $S_{11.3}/S_{9.8}=1.0$, we find that 13 sources in our sample satisfy this criterion, and thus require emission near 11.3~$\mu$m that is in excess above models of amorphous olivines/pyroxenes, indicating that the features arise from crystalline silicates or PAHs.   These sources are marked with a ``Y'' in the column 3 of Table~\ref{tab:features}.
Of these 13 sources, 4 show clear evidence of crystalline silicate emission at $\lambda > 25$~$\mu$m (DoAr 24E, GY 23, HD 98922 and DL Cha).  
Clear evidence for PAH emission features in the SL spectra can be seen in 3 additional sources (LkH$\alpha$ 330, IRAS 03446+3254, Haro 1-17 and HD 135344). The spectra of 3 more sources (VSSG1, LkH$\alpha$ 327, and RR Tau) show evidence of both PAH and crystalline silicate features. The remaining 3 sources with $S_{11.3}/S_{9.8}>1.0$ (T Cha, SR 21, and HD 135344) display narrow isolated 11.3~$\mu$m emission and have very simple spectra that show only shallow $\sim$20~$\mu$m silicate emission.  This likely indicates that the 11.3~$\mu$m features in these 3 spectra arise from PAH, and not crystalline silicate, emission.  Therefore, for all of the spectra with $S_{11.3}/S_{9.8}>1.0$, the additional flux at 11.3~$\mu$m can indeed be explained by crystalline silicate and/or PAH emission.  
The nature of the 11.3~$\mu$m features and relationship to PAH emission will be discussed more thoroughly in Geers et al. (in prep.). 
Aside from the spectra with $S_{11.3}/S_{9.8}>1.0$, the data presented in Figure~\ref{fig:vb}a are consistent with grain size variations.

The 20~$\mu$m features show a shape-strength trend which is very similar to that seen for the 10~$\mu$m features (Figure~\ref{fig:vb}b). Here we use the ratio of the flux at 23.8 to 19~$\mu$m, again the peaks of the most prominent amorphous and crystalline silicate features, as a proxy for the feature shape.  This plot includes only c2d data, as continuum normalized ISO data were not available.  The paucity of spectra with strong 20~$\mu$m features may be related to the prominence of weak, flat 10~$\mu$m features in the c2d sample, indicating large grain sizes.  Although the sample size (and range in peak strength) is smaller for the 20~$\mu$m features, the shape-strength trend is still prominent.  This trend possesses a similar slope and y-intercept to that seen for the 10~$\mu$m features and is again most closely matched with the Mie model of an olivine/pyroxene mixture, which agrees to within $S_{23.8}/S_{19}=0.01$, or $\sim$1\%.  There appears to be little difference in grain composition between grains being probed by the 10 and 20 $\mu$m features.  

All of the 20 $\mu$m features are consistent with  models of amorphous olivine/pyroxene mixtures of sizes between 0.1 and 20 $\mu$m, with no deviations indicative of strong crystalline features.  A similar fit is obtained for the Mie and DHS models of the olivine/pyroxene mixture, but for the DHS models much larger grain sizes ($a>20$~$\mu$m) are needed.  The data are also consistent with DHS and Mie models of amorphous olivines (to within $S_{23.8}/S_{19}=0.08$, or $\sim$8\%), for much smaller grain sizes ($a{\le}5$~$\mu$m).  This makes exact grain-size determination difficult.   Furthermore, a large scatter in the modeled data points can be seen for $S^{20{\mu}m}_{peak}<1.1$ due to difficulties in fitting the continuum for such weak features.  This prevents identification of small deviations in feature shape of between the models and data for large grain sizes (as were seen for the 10~$\mu$m features).  However, larger deviations from the models are not seen.  Thus, although there is evidence in some individual spectra for crystalline silicate features, the entire trend is consistent with amorphous olivine/pyroxene mixtures with source-to-source size variations.  

For the 48 sources observed in this study, the 10 and 20~$\mu$m features both exhibit strength-shape trends consistent with source-to-source grain-size variations, with most of the observed features similar to models of grains with sizes much larger than that of the ISM ($a>>0.1\mu$m).  This indicates that grain growth must occur quickly in these disks. 
Furthermore, as the observed features arise from silicate emission in the disk surface layers, and the largest grains will gravitationally settle to the disk midplane, the emitting grains likely represent the low-size tail of the silicate grain size distribution within the disk.  This effect is enhanced by the fact that larger grains have weaker silicate emission features.

\subsection{10 vs. 20~$\mu$m feature strengths}

\begin{figure*}
\includegraphics[angle=90,width=7in]{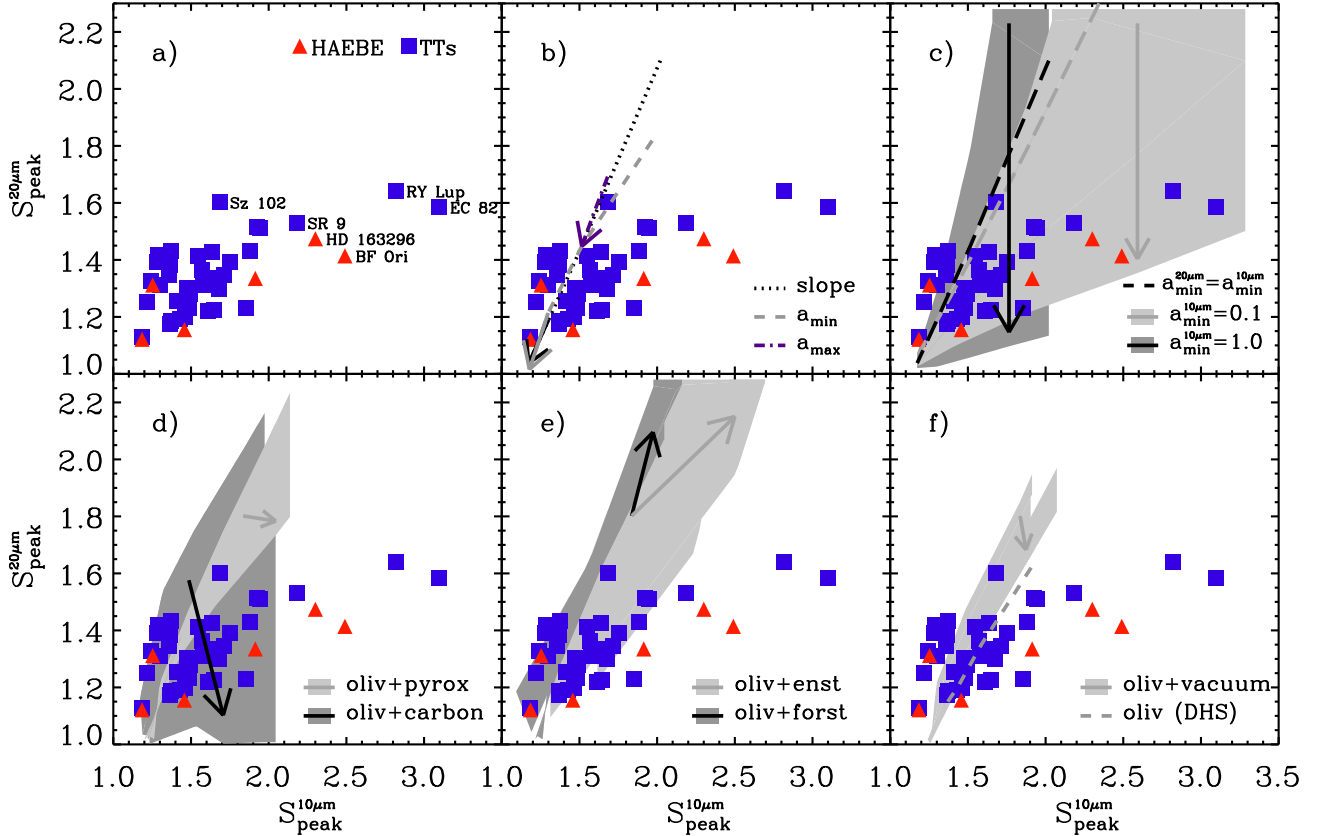}
\figcaption{Relative strengths of the 20 versus 10~$\mu$m features. Models are overplotted showing variations in $b)$ grain size distribution, $c)$ ratios of grain sizes emitting at 10 and 20~$\mu$m, $d)$ grain composition, $e)$ crystallinity, and $f)$ porosity.  The reference opacities are calculated via Mie theory for pure olivine grains with differential grain size distributions with $a_{min}=1.0 \mu$m, $a_{max}=100 \mu$m, and power-law index $p=-3.5$ for both the 10 and 20~$\mu$m features. Arrows in panel $b$ indicate the direction of increasing slope $p$, $a_{min}$ and $a_{max}$. Gray shaded regions in $c)$ indicate the range in 10 and 20~$\mu$m feature strengths covered by variations in $a_{min}^{20{\mu}m}$ with  $a_{min}^{10{\mu}m}=0.1$~$\mu$m and $a_{min}^{10{\mu}m}=1.0$~$\mu$m.  Arrows in panel $c$ indicate the direction of increasing 10-to-20~$\mu$m grain size ratios over the gray shaded regions.  Gray shaded regions in $d$--$f$ indicate the range in 10 and 20~$\mu$m feature strengths covered by variations in x/olivine ratios, where x is $d)$ pyroxene or amorphous carbon, $e)$ forsterite or enstatite, and $f)$ vacuum for grains of constant size. Arrows in $d$--$f$ indicate the direction of increasing x/olivine ratios.  Symbols are defined as in Figure~\ref{fig:vb}.
\label{fig:f10vs20}}
\end{figure*}

\begin{deluxetable}{lccc}
\tablecaption{Model parameters for Figure~\ref{fig:f10vs20}. \label{tab:10vs20}}
\tablehead{\colhead{Panel} & \colhead{Parameter} & \colhead{Variables} & \colhead{Range}}
\startdata
b               &size distribution\tablenotemark{a}  & $a_{min}$  & 0.01--3.0~$\mu$m \\
                &                      & $a_{max}$                        & 10--100~$\mu$m \\
                &                      & $p$                              & 4.5--2.5 \\
\tableline 
c               & sizes for 10 vs 20~$\mu$m & $p$                         & 4.5--2.5 \\
                &                      & $a_{min}^{10}$                   & 0.1,1.0~$\mu$m \\
                &                      & $a_{min}^{20}$                   & (0.1--20)$\times$0.1~$\mu$m \\
                &                      & $a_{min}^{20}$                   & (0.5--3)$\times$1.0~$\mu$m \\
\tableline
d               & composition          & $a_{min}^{10}$ = $a_{min}^{20}$  & 0.01--3.0~$\mu$m \\
                &                      & olivine:carbon                   & 100:0--50:50 \\
                &                      & olivine:pyroxene                 & 100:0--0:100 \\
\tableline
e               & crystallinity        & $a_{min}^{10}$ = $a_{min}^{20}$  & 0.01--3.0~$\mu$m \\
                &                      & olivine:forsterite               & 100:0--0:100 \\
                &                      & olivine:enstatite                & 100:0--0:100 \\
\tableline
f               & porosity             & $a_{min}^{10}$ = $a_{min}^{20}$  & 0.01--3.0~$\mu$m \\
                &                      & olivine:vacuum                   & 100:0--20:80 \\
\enddata
\tablenotetext{a}{Using a differential grain size distribution of $dn(a){\propto}a^{-p}da$ where the absorption efficiency is defined as $<Q_{abs}>=\int_{a_{min}}^{a_{max}} Q_{abs} dn(a) / \int_{a_{min}}^{a_{max}} dn(a)$.}
\end{deluxetable}

One of the most interesting aspects of this sample is a set of sources with very strong $S^{10{\mu}m}_{peak}$ but weak $S^{20{\mu}m}_{peak}$. 
These ``outliers'' can be clearly seen in a plot of the strengths of the 10 and 20~$\mu$m features in the observed spectra (Figure~\ref{fig:f10vs20}; Table~\ref{tab:f10vs20}). The feature strengths appear to be correlated over the entire sample ($r=0.36$, 96\% probability of correlation). 
However, examination of Figure~\ref{fig:f10vs20}a reveals that the bulk of the data is clustered around $S^{20{\mu}m}_{peak}\approx1.3$ and $S^{10{\mu}m}_{peak}\approx1.5$ and $\sim$11 outliers have 10~$\mu$m and 20~$\mu$m features with very different strengths.  Furthermore, most of these outliers have strong 10~$\mu$m features ($S^{10{\mu}m}_{peak}>2$) and weak 20~$\mu$m features ($S^{20{\mu}m}_{peak}<1.6$). No sources in our sample have strong 10~$\mu$m features {\it and} equally strong 20~$\mu$m features.  One may expect weak 20~$\mu$m features to be more abundant as most sources show weak 10~$\mu$m features (cf. Figure~\ref{fig:vb}a) and grain growth is occuring quickly in these disks.  However, the fact that there are {\it no} sources with strong 20~$\mu$m features and weak 10~$\mu$m features is significant.  

In order to interpret the distribution of peak strengths in the observed spectra, we examine the effect of a variety of grain parameters on the strengths of the 10 and 20~$\mu$m features.  Factors affecting silicate feature strengths include, but are not limited to, the grain size distribution, relative sizes of grains emitting at 10 and 20~$\mu$m, grain composition, crystallinity, and porosity.  We attempt to assess the influence of each of these parameters in panels b--f of Figure~\ref{fig:f10vs20} and $\S$5.2.1--5.2.4 for models of normalized $Q_{abs}$ calculated as described in $\S$4.2.  

\subsubsection{Grain size distributions}

The effects of grain size are explored by modeling opacities for a set of grains with a differential grain size distribution $dn(a) \propto a^{-p}\; da$, and varying the minimum grain size $a_{min}$, maximum grain size $a_{max}$, and power-law index $p$, as shown in Figure~\ref{fig:f10vs20}b. All three parameters have very similar effects; increasing $a_{min}$ or $a_{max}$ or decreasing $p$ results in more large grains and reduces both the 10 and 20~$\mu$m feature strengths, but does not affect the ratio of the two.  Thus, variation of the grain size distribution cannot account for the extreme 10-to-20~$\mu$m ratios of the outliers in Figure~\ref{fig:f10vs20}a. The bulk of the observed data set is centered around 10 and 20~$\mu$m feature strengths similar to that of a grain size distribution with $a_{min}=1.0$~$\mu$m, $a_{max}=100$~$\mu$m, and power-law index $p=3.5$. 

\begin{figure}
\includegraphics[angle=0,width=8cm]{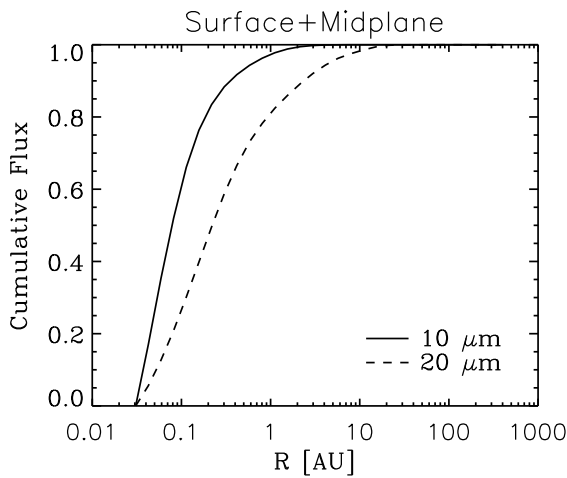}
\figcaption{Cumulative flux at 10~$\mu$m (solid line) and at 20~$\mu$m (dashed line) as a function of radius. Fluxes are calculated using the CGPLUS model \citep{DDN01a} for a disk around a T~Tauri star with $T_{\star}=4000$~K, $L_{\star}=0.58$~$L_{\odot}$ containing olivine grains of a differential grain size distribution with $a_{min}=1$~$\mu$m, $a_{max}=100$~$\mu$m and $p=-3.5$. 95\% of the 10~$\mu$m flux comes from within $\sim$1~AU, while 95\% of the 20~$\mu$m flux comes from within $\sim$10~AU.
\label{fig:cumflux}}
\end{figure}

The emission at 10 and 20~$\mu$m likely comes from different populations of grains.  This is a simple consequence of the temperature dependence of the emission as a function of wavelength (e.g., emission at 20~$\mu$m can arise from colder grains than emission at 10~$\mu$m).  We therefore expect the contribution of a particular grain to the 10 and 20~$\mu$m features to depend on the location of the grain.
In order to determine the relative contribution of grains from different radii to the 10 and 20~$\mu$m features, we perform an exercise using a simple two-layer model \citep[CGPLUS,][]{DDN01a,CG97} to calculate the cumulative disk flux as a function of radius at 10 and 20~$\mu$m for a typical T~Tauri star. Figure~\ref{fig:cumflux} shows that the radii probed by the 10 and 20~$\mu$m silicate features can indeed be quite different, with the 20~$\mu$m feature probing radii up to 10 times those probed by the 10~$\mu$m feature.  

In Figure~\ref{fig:f10vs20}c, the relationship between the sizes of the grains emitting at 10 and 20~$\mu$m is explored.  We vary $a_{min}$ for the models of the 20~$\mu$m feature and keeping $a_{max}$, $p$, and $a_{min}$ for the models of the 10~$\mu$m feature constant.  The entire distribution of the data can be reproduced by these models. Most data correspond to $a_{min}^{10{\mu}m}=1$~$\mu$m and $a_{min}^{20{\mu}m}=0.1$--3~$\mu$m, but the data in the right side of Figure~\ref{fig:f10vs20}c require large differences between the sizes of grains emitting at 10 and 20~$\mu$m ($a_{min}^{10{\mu}m}=0.1$~$\mu$m, $a_{min}^{20{\mu}m}=2$--3~$\mu$m).  

The 10 and 20~$\mu$m features are clearly probing different dust populations.
The bulk of the data can be reproduced by models of 1.0~$\mu$m grains, indicating significant growth in these disks compared to the ISM.   Although the data are clustered centered around the dashed lines that denote equal grain sizes for 10 and 20~$\mu$m emission, many spectra can be better represented by much larger grains emitting at 20~$\mu$m than at 10~$\mu$m.  This may be explained, if the emission is optically thin, by a difference in the $\tau=1$ surface depth as a function of wavelength of the emission; 20~$\mu$m features probe a deeper layer of the disk where grain sizes are likely larger due to larger density and dust settling.  One would also expect, however, that dust at larger radii (and lower temperature), which is likely smaller due to decreased densities, contributes more to the 20~$\mu$m feature than the 10~$\mu$m feature.  
There is thus a competing effect between the disk height and radius being probed. This may explain why most of the data points are centered around the dashed lines that denote equal grain sizes for 10 and 20~$\mu$m emission.
In the case of the outliers, enhanced dust settling may make the disk height differential between dust probed by the two features more important than the difference in radius.

\subsubsection{Grain composition and crystallinity}

Grain composition and crystallinity also affect the strengths of the silicate emission features, resulting in ranges of peak strengths that are very similar to those produced by varying the grain sizes (with $a_{min}^{10{\mu}m}=a_{min}^{20{\mu}m}$).  Thus, effects of grain-size variation, composition and crystallinity are difficult to disentangle using 10-to-20~$\mu$m flux ratios.

In Figure~\ref{fig:f10vs20}d, we vary the percentage of amorphous pyroxene and amorphous carbon relative to amorphous olivine.
To simplify matters, we use the same composition and grain sizes for the grains emitting at 10 and 20~$\mu$m.  Increasing the pyroxene-to-olivine fraction has the largest effects for small grains ($<$1.0~$\mu$m), resulting in an increased 10~$\mu$m feature strength and roughly constant 20~$\mu$m feature strength (as seen previously in Figure~\ref{fig:vb}).  Increasing the carbon-to-olivine fraction has an even larger effect on grains of small sizes, with a larger effect on the 20~$\mu$m feature than the 10~$\mu$m feature, as the peak-to-continuum ratio is decreased due to an increase in the effective ``continuum.'' 

Variations in crystallinity (using DHS models) also result in the same range of peak strengths, but highly crystalline grains can produce features with the same peak strengths at smaller grain sizes (Figure~\ref{fig:f10vs20}e).  When the fraction of forsterite or enstatite is less than $\sim$50\%, the feature strengths remain very similar to amorphous olivine. Beyond a crystalline fraction of $\sim$50\%, the strong and narrow crystalline emission features begin to dominate and the strengths of the peak emission in the 10 and 20~$\mu$m regions appear to increase dramatically for the same grain size distribution.  For models of $>$50\% crystalline forsterite and enstatite, the gray shading in Figure~\ref{fig:f10vs20}e extends far beyond the plotted range of the y-axis ($S^{20{\mu}m}_{peak}>>2.28$).  Thus, for primarily crystalline grains, grain size distributions with $a_{min}\le2.0$~$\mu$m can account for the entire gray area shown.

Differing composition and crystallinity can account for the 10 and 20~$\mu$m ratios of the bulk of the observed spectra, but not the outliers. 
 Additionally, the 10 and 20~$\mu$m ratios of models of varying grain composition and crystallinity encompass ranges in which no observational data lie. In particular, spectra showing weak 10~$\mu$m features ($S^{10{\mu}m}_{peak}>1.7$) {\it and} strong 20~$\mu$m features can be easily produced with models of small-sized ($a_{min}<1.0$~$\mu$m) amorphous silicate/carbonaceous grains or moderately-sized ($a_{min}<3.0$~$\mu$m) crystalline silicate grains, but they are not seen in the $c2d$ sample.  These absences confirm the idea that the 20~$\mu$m feature probes regions of larger grain size than does the 10~$\mu$m feature, irrespective of grain composition and crystallinity.

\subsubsection{Dust porosity}

Finally, in Figure~\ref{fig:f10vs20}f we explore the effects of grain opacity by increasing the vacuum fraction of a set of olivine grains and by using the distribution of hollow spheres method (DHS).  In both cases, we integrate over a grain size distribution as in Figure~\ref{fig:f10vs20}d--e. The primary difference between the two methods is that the opacities calculated with DHS are the average of a set of hollow spheres of varying vacuum fractions for a given mass, which is translated to grain size. The $Q_{abs}$ are then integrated over a grain size distribution. For the olivine+vacuum grains, the opacities and $Q_{abs}$ are calculated and integrated over grain size for each volume fraction individually. The variation over vacuum fraction (light gray in Figure~\ref{fig:f10vs20}f) does not have a large effect on the 10-to-20~$\mu$m ratio and is consistent with the DHS 10 and 20~$\mu$m strengths and variation with $a_{min}$.  Variations in dust porosity can only explain a fraction of the bulk of the sample, which overlaps with the regions covered by models of varying grain composition, and cannot explain the outliers.  Thus it is not necessary to consider porosity to explain the observed 10 and 20~$\mu$m feature strengths.

\section{Grain growth and stellar/disk properties}

\begin{figure*}
\includegraphics[angle=90,width=7in]{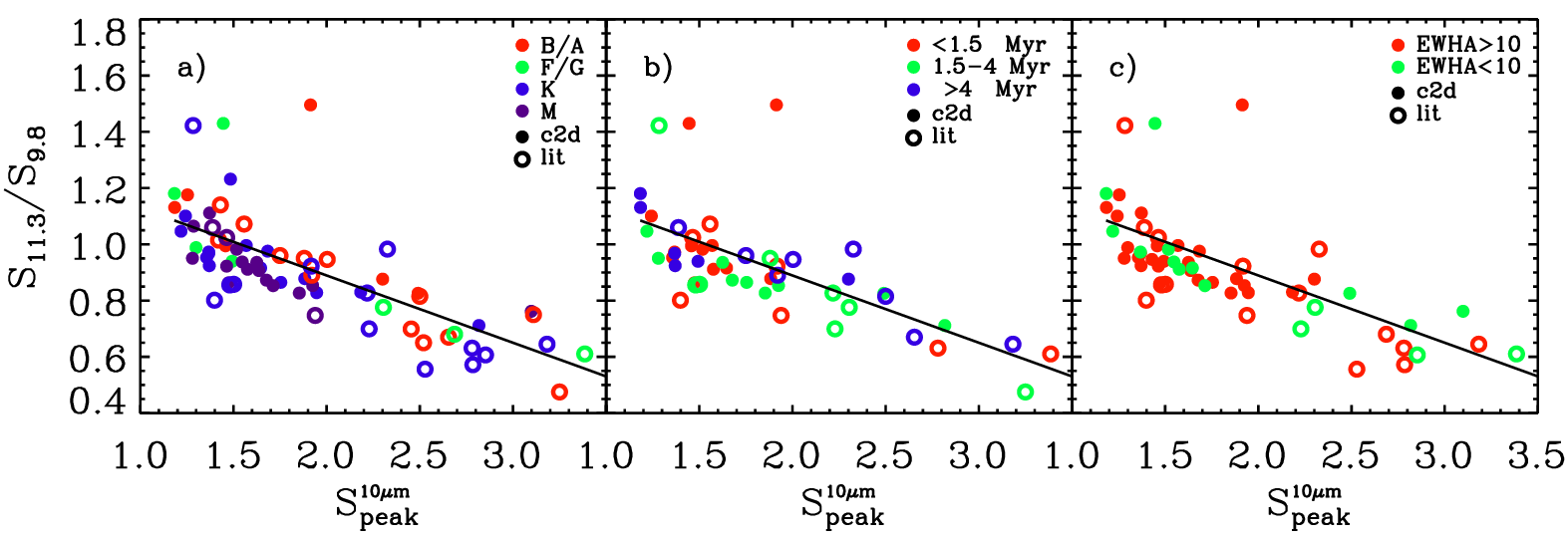}
\figcaption{Shape and strength of the 10~$\mu$m silicate features. The shape of the feature ($S_{11.3}/S_{9.8}$) is plotted versus the feature strength ($S^{10{\mu}m}_{peak}$) in all panels.  The points are color-coded by $a)$ spectral type, $b)$ age and $c)$ H$\alpha$ equivalent width. Silicate features from the literature \citep{vanboekel_03, przygodda_03,me_05} are always represented by open symbols. The black line in all panels indicates the correlation found for the 10~$\mu$m features shown in Figure~\ref{fig:vb}a including all observations. This trend is not strongly dependent on H$\alpha$EW or age, but may be related to spectral type.
\label{fig:vbprop}}
\end{figure*} 

In order to evaluate the dependence of the strength-shape trend noted in $\S$5 on stellar and disk parameters, we plot the 10~$\mu$m shape versus strength again in Figures~\ref{fig:vbprop}a--c, color-coded by spectral type, stellar age, and H$\alpha$ equivalent width (for TTs only). Stellar/disk parameters for each source observed in this study ({\it c2d}) were collected from the literature and are shown in Table~\ref{tab:obs}. Stellar/disk parameters for 10~$\mu$m data obtained from \citet{vanboekel_03}, \citet{przygodda_03}, and \citet{me_05} ({\it lit}) are shown in Table~\ref{tab:lit}.  
 We evaluate the probability that the sets of spectral type, stellar age or H$\alpha$ equivalent width bins are drawn from the same distribution on the 10~$\mu$m feature shape-strength relation shown in Figure~\ref{fig:vb}a by evaluating the two-sided Kolmogorov-Smirnov statistic ($D$)\footnotemark~ for each stellar/disk parameter.  As we are evaluating the dependence of the strength-shape trend on stellar/disk parameters, we use only the sources with $S_{11.3}/S_{9.8}<1.0$, thus removing sources with substantial emission from crystalline silicates and/or PAHs.  Some stellar/disk parameters are not available in the literature, and therefore not all sources are included in Figures~\ref{fig:vbprop}a, b, and c.
\footnotetext{The two-sided Kolmogorov-Smirnov statistic is the difference between the cumulative distribution function (CDF) of two sets of data and is calculated using {\bf KSTWO} \citep[from Numerical Recipes;][]{Numerical}.}

\subsection{Spectral Type}

\begin{figure}
\includegraphics[angle=90,width=8cm]{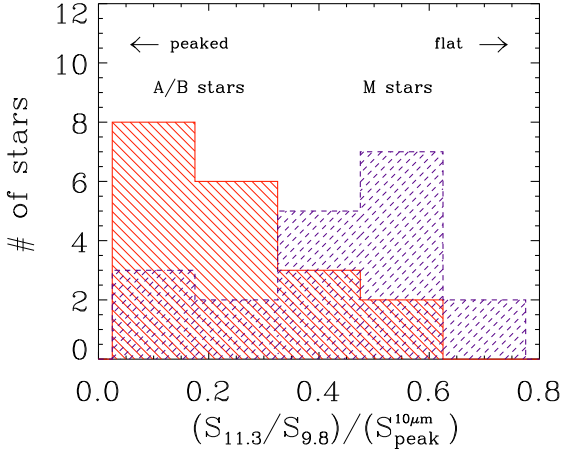}
\figcaption{Histogram of the 10~$\mu$m feature shape-to-strength ratios for the A/B and M stars shown in Figure~\ref{fig:vbprop}a.  Spectra of A and B stars show more centrally peaked 10~$\mu$m features, lying in the middle and left of Figure~\ref{fig:vbprop}a, while spectra of M stars show primarily flattened features, lying in the upper left of Figure~\ref{fig:vbprop}a. The bin widths are 0.1 in units of ($S_{11.3}/S_{9.8}$)~/~($S^{10{\mu}m}_{peak}$).
\label{fig:histsptype}}
\end{figure}

As the data in our sample (filled symbols) appear to be consistent with the largest silicate grain sizes, and probe disks around primarily K and M stars, we first looked for connections between grain size and spectral type.  We divide the sample by spectral type into four groups: 1) A/B, 2) F/G, 3) K and 4) M. The A and B stars and F and G stars were combined into single groups as few stars of each type were observed.
The Kolmogorov-Smirnov (K-S) test is used to compare two arrays of data values and the process is repeated for each combination of groups 1 through 4. In general, there appears to be a dependence on spectral type, with A/B stars being clustered toward the middle and lower right of the 10~$\mu$m feature strength-shape trend and M stars being more clustered toward the upper left of Figure\ref{fig:vbprop}.  This can be more easily seen in Figure~\ref{fig:histsptype}, where histograms of the 10~$\mu$m shape-to-strength ratio for A/B stars and M stars are compared.  The population of A/B stars peaks at ratios near 0.1 and decreases for larger ratios, while only a few M stars have shape-to-strength ratios near 0.1 and the population increases for larger ratios to peak near 0.5--0.6.

Although it appears that 10~$\mu$m silicate emission may be different in A/B stars versus M stars, a larger data set is needed to confirm this result.
The calculated probabilities that these groups are drawn from the same distribution range from 15\% ($D=0.29$) for spectral types A/B and M to 98\% ($D=0.17$) for spectral types A/B and F/G.  Objects that are close in spectral type, in subsequent groups (e.g., 1 vs. 2, 2. vs. 3), have large probabilities (69--98\%) of being drawn from the same distribution. Groups with larger differences in spectral type (e.g., 1 vs. 3, 2. vs. 4) have smaller probabilities (15--50\%).  
Of all these comparisons, however, only the K-S test for group 1 vs. 4 (A/B vs. M) indicates a low enough probability \citep[$\le20$\%,][]{Numerical} to suggest that the differences in the populations are statistically significant.  Furthermore, the number of objects in each group is still quite low (13 A/B stars, 4 F/G stars, 24 K stars, and 15 M stars) and  the spectral types typically have errors on the order of 3 subclasses.

The dependence of the strength-shape trend with spectral type may be explained in terms of the disk temperature and density structure.  If we assume that the 10~$\mu$m feature always probes grains of roughly the same temperature, then the radius being probed by the 10~$\mu$m emission will change as a function of the temperature structure of the disk.  The disk temperature $T(R)$ varies as a function of radius as 
$$T(R)={\alpha}^{1/4} \sqrt{\frac{R_{\star}}{R}} T_{\star},$$ 
\citep[from][]{DD05}, where $\alpha$ is the disk flaring angle and $R_{\star}$ and $T_{\star}$ are the stellar radius and temperature. $R$ is the radius corresponding to temperature $T(R)$, in this case the dominant radius being probed by the silicate emission feature, which we will call $R_{sil}$.  Assuming that $T(R_{sil})$ and $\alpha$ are constant over the entire disk sample,
and using typical values of $R_{\star}$ and $T_{\star}$ for A/B and M stars, $R_{sil,A/B}/R_{sil,M}{\approx10}$--$25$.  So the
radius probed by the 10~$\mu$m emission feature is quite different for stars of spectral types A/B and M.  Furthermore, the density generally increases with decreasing radius and grain growth increases at higher density, while dust settling decreases at higher density. Thus, the combination of faster grain growth and slower dust settling at smaller radii could lead to the larger grains that we see in M stars versus A/B stars.  Although this simple theory is consistent with the observed correlation of grain size with spectral type, several other spectral-type dependent factors (e.g., X-rays, UV radiation, stellar/disk winds, etc.) likely influence grain size in these disks.

\subsection{Age}

Although one might expect grain growth and crystallization to occur over time, as the disk matures, we see no clear relationship between the strength-shape trend and stellar age (Figure~\ref{fig:vbprop}b). 
The ages for our sample are culled from the literature and range from 0.5 to 6 Myr with errors of up to a few Myr \citep[i.e.,][]{HW04}.  We therefore divide up the sample into three approximately equally sized groups: stars with ages of 1) $<$ 1.5 Myr, 2) 1.5--4 Myr, and 3) $>$ 4 Myr, and follow a similar method as described above for the analysis of spectral type.  
The strength-shape trend does not appear to be strongly related to the age of the star (Figure~\ref{fig:vbprop}b), with probabilities of $>$47\% that all three groups are drawn from the same distribution.
However, this analysis would be more conclusive for stars within one cluster with a well defined age, for which relative ages could be more accurately determined.  With the available ages, these results indicate that the stellar age of PMS stars is not directly related to the state of the small, $\mu$m-sized grains in these disks.

In addition to the difficulties in determining the disk age, the lack of an age dependence of the shape-strength trend can be understood when grain fragmentation and disk turbulence are considered.  \citet{DD05} find that in models including only basic coagulation mechanisms grain growth occurs very quickly, such that the SEDs show much weaker IR excess (even at young ages of 1 Myr) than do disks around most classical TTs. The settling of small grains, removing them from the disk surface, occurs over even faster time scales.  The lack of sources with 
strongly peaked 10~$\mu$m features (and therefore small grains) in the substantial sample observed here is consistent with this fast timescale for grain growth (to 1~$\mu$m sizes).  The presence of strongly peaked 10~$\mu$m features  in some sources (particularly in the $lit$ sample) may be indicative of the replenishment of small grains in disk surface layers through mechanisms such as fragmentation and/or turbulent mixing.  (Some amount of large grains will also be mixed into disk surface layers).  The models presented in \citet{DD05} suggest that such mechanisms may allow disk systems to maintain quasi-stationary grain size distributions, with grain growth and small-grain replenishment rates in equilibrium for long periods of time ($\ge$1 Myr).  Thus the typical grain sizes found in disks around pre-main sequence stars are likely not to be directly related to disk age, but indicative of a number of other factors, such as the strength of the turbulence and the gas mass in the disk.

\subsection{Accretion}

We also do not see a correlation of the strength-shape trend with the equivalent width of the H$\alpha$ emission lines (H$\alpha$EW), probing disk evolution (Figure ~\ref{fig:vbprop}c). The H$\alpha$ equivalent width is a tracer of active stellar accretion from the inner disk. Thus clearing of the inner disk will result in lower H$\alpha$EW.   In Figure ~\ref{fig:vbprop}c, we separate the  TTs in our sample into two groups using the equivalent width of H$\alpha$: 35 classical T~Tauri stars  (cTTs; H${\alpha}$EW~$>10$~$\rm\AA$) and 14 weak-lined T~Tauri stars (wTTs; H${\alpha}$EW~$<10$~$\rm\AA$).  Although there is no clear observational evidence that cTTs are progenitors to wTTs, wTTs are often described as cTTs in which the inner disk has dissipated.  Therefore, we might expect that the clearing of the inner disk, and thus the transition between the classical and weak-line T~Tauri star phases, would be related to grain growth and the 10~$\mu$m feature strength-shape trend.  We find that there is a high probability ($\sim$70\%) that the two groups are drawn from the same distribution, however, indicating that there is no clear relationship between H$\alpha$EW and grain growth in these disks.  This indicates that either H$\alpha$EW is not a good probe of the disk evolutionary state or stellar accretion rate or that grain growth is not related to these quantities.  We also note that H${\alpha}$ emission can be quite variable and the values used here are time averages, using the full range of observed H${\alpha}$EW as error bars.  The relation between H${\alpha}$EW and silicate feature shape-strength would more precisely tested with simultaneous MIR and optical spectroscopy.

\section{Conclusions}

Spectra in the $\sim$5--35~$\mu$m region have been obtained for disks around 40 TTs and 7 HAEBE stars using the Spitzer Space Telescope, as a subset of the c2d IRS survey.  This is the first significant sample of T~Tauri stars for which both 10 and 20~$\mu$m features are available.
Broad silicate features at 10 and 20~$\mu$m,  similar to emission from amorphous silicates, are prominent toward the observed sample of TTs and HAEBE stars.  Emission features from crystalline silicates are also evident in the observed spectra, with the most prominent features visible in the 33-36~$\mu$m region.  

We performed a statistical analysis of the shapes and strengths of the 10 and 20~$\mu$m features and find the following: 
\begin{list}{---}
\item{If the full IRS spectra are considered, the data are
most consistent with source-to-source variations in grain size, with
the bulk of the sources in our sample indicating sizes of 1~$\mu$m or
greater. This and the lack of strongly peaked 10~$\mu$m sources in our sample of 40 sources is consistent with fast grain growth (from 0.1--1.0~$\mu$m sizes) in the surfaces of these disks.}
\item{}
\item{Long wavelength (33-36~$\mu$m) crystalline silicate features are seen toward $\sim$half of the TTs disks in our sample.  This indicates that significant dust processing is also occuring in these disks.  Only 13 of these spectra also exhibit 11.3-to-9.8~$\mu$m ratios that cannot be reproduced by models of amorphous silicates, suggesting that the degree of crystallinity deduced from the 10~$\mu$m region alone is underestimated.}
\item{A subset of the observed spectra have particularly large 10-to-20~$\mu$m ratios, which can only be reproduced with much smaller grains emitting at 10~$\mu$m ($a_{min}=0.1$~$\mu$m) than at 20~$\mu$m ($a_{min}=2-3$~$\mu$m). This can be explained if the 20~$\mu$m emission arises from deeper in the disk than the 10~$\mu$m emission in disks where dust settling has occurred.}
\item{The 10~$\mu$m feature strength vs. shape trend {\it is not} correlated with age or disk evolutionary state (H$\alpha$EW).  This suggests the importance of turbulence and the regeneration of small ($\mu$m-sized) grains on the disk surface.}
\item{The 10~$\mu$m feature strength vs. shape trend {\it is} related to spectral type, with M stars showing significantly flatter silicate features (larger grain sizes) than those of A/B stars.  This may be related to a difference in the radius probed by the emission, which should increase as a function of the disk temperature, and thus stellar luminosity. However, the observed correlation could also be indicative of other spectral-type dependent factors (e.g, X-rays, UV radiation, stellar/disk winds, etc.).}
\end{list}

As the observed features arise from silicate emission in the disk surface layers, and the largest grains will gravitationally settle to the disk midplane, the emitting grains likely represent the small-size tail of the silicate grain size distribution within the disk. This simple picture of grain growth and settling is complicated by vertical mixing, which can bring both large and small grains back to the disk surface, and fragmentation, which results in the replenishment of small grains throughout the disk.  Comparisons with probes of grain size in the disk midplane and inner disk clearing, as well as probes of disk turbulence and gas-dust interactions, are vital to understanding the results presented here in the context of planet formation.


\acknowledgments

Support for this work, part of the Spitzer Space Telescope Legacy
Science Program, was provided by NASA through Contract Numbers 1256316, 
1224608 and 1230780 issued by the Jet Propulsion Laboratory, California
Institute of Technology under NASA contract 1407.  
Astrochemistry in Leiden is supported by a NWO Spinoza 
and NOVA grant, and by the European Research Training Network "The Origin of
Planetary Systems" (PLANETS, contract number HPRN-CT-2002-00308).
The authors would like to thank Michiel Min for communicating his results on DHS and Jeroen Bouwman, Thomas Henning, Roy van Boekel, Rens Waters, Antonella Natta, David Koerner and the ApJ referee for helpful comments and suggestions.

\bibliographystyle{apj}


\end{document}